\documentclass[
superscriptaddress,amsmath,amssymb,aps,twocolumn,
notitlepage
]{revtex4-2}

\usepackage{graphicx}% Include figure files
\usepackage{dcolumn}% Align table columns on decimal point
\usepackage{bm}% bold math
\usepackage{lipsum}
\usepackage[dvipsnames]{xcolor}
\usepackage[T1]{fontenc}
 \fontfamily{cmr}

\usepackage{float}
\usepackage{hyperref}
\usepackage{titlesec}
\usepackage{lipsum}
\usepackage{tikz-cd}
\usepackage{todonotes}
 
\tikzcdset{every label/.append style = {font = \small}}

\titleformat*{\section}{\large\bfseries}
\usepackage[normalem]{ulem}
\hypersetup{
    colorlinks=true,
    linkcolor=blue,
    citecolor=magenta,      
    urlcolor=cyan,
    }

\begin{document}

\preprint{APS/123-QED}

%\title{Laser-written modulations of structured light integrated in glass}
\title{Volumetric Processing of Structured Light Integrated in Glass}
%\title{Toolbox for advanced modulations of structured light in glass}
%framework 
% Volumetric Structured Light Processing Integrated in Glass
% Volumetric Processing Structured Light Integrated in Glass

\author{O. Korichi}
\email{oussama.korichi@tuni.fi}

\affiliation{Tampere University, Photonics Laboratory, Physics Unit, Tampere, FI-33720, Finland}

\author{M. Hiekkamäki}%
\affiliation{Tampere University, Photonics Laboratory, Physics Unit, Tampere, FI-33720, Finland}
\affiliation{Quantum Technology Laboratories GmbH, Clemens-Holzmeister-straße 6/6, 1100 Vienna, Austria}
\author{R. Fickler}
\affiliation{Tampere University, Photonics Laboratory, Physics Unit, Tampere, FI-33720, Finland}

\date{\today}% It is always \today, today,

%info on Nature Photonics ARTICLE
%Main text – up to 3,000 words, excluding abstract, Methods, references and figure legends.
%Abstract – up to 200 words, unreferenced. 
%Display items – up to 6 items (figures and/or tables). 
%Article should be divided as follows: 
%    Introduction (without heading) 
%    Results
%    Discussion
%    Online Methods.
%Results and Methods should be divided by topical subheadings; the Discussion does not contain subheadings.
%References – as a guideline, we typically recommend up to 50.
%Articles include received/accepted dates. 
%Articles may be accompanied by supplementary information. 
%Articles are peer reviewed.

\begin{abstract}

Light with complex structures in polarization, phase and amplitude, has attracted a lot of attention in a broad range of applications and fundamental studies in classical and quantum optics.  
Along with the increased interest in structured light comes a need for efficient modulation platforms operating simultaneously for many modes. 
Multi-plane light conversions (MPLC), i.e., multiple consecutive phase modulations in combination with free space propagation, have enabled such unitary transformations, which are usually built by bulky optical components, limited to scalar modulation, or rely on advanced nanofabrication techniques. 
Here, we demonstrate an efficient, monolithic MPLC architecture through direct laser writing in standard fused silica glass, resulting in a device with a compact form factor of only a few mm$^3$.
Our scheme is based on volumetric engineering of the glass's birefringence through laser-written nanogratings, which enables spatial control over full vectorial light structures. 
To showcase the approach's potential for integrated multimode–multipath optical networks, we demonstrate multi-mode unitary transformations, mode conversions, and complex beam-splitting for scalar light.
We further extend the MPLC operation to vectorial light and implement various polarization-controlled spatial mode operations as well as the transformation of the topology of an optical Skyrmion.
Finally, we highlight our scheme's promise for optical communications and implement a miniaturized multiplexer for spatial modes and polarization operating at telecom wavelength.

\end{abstract}

\maketitle

%%%% MAIN TEXT - 3000 words

% \noindent \textbf{Introduction}\\

\section*{\textbf{Introduction}}
Engineering light in all its degrees of freedom (DOF), i.e., the transverse complex amplitude, polarization, and the temporal or spectral domain, has evolved from modulations of a single DOF to more generalized approaches encompassing simultaneous control over multiple domains. 
Such efforts are commonly summarized under the field of \textit{structured light} \cite{he2022towards}%piccardo2022roadmap, shen2023roadmap}
, of which a particularly prominent section focuses on complex shaping in the spatial domain \cite{forbes2021structured}. 
%A seminal development in this direction was the realization that light can exhibit orbital angular momentum when its wavefront is structured to form a helix around the optical axis of propagation \cite{allen1992orbital}.\par
The concept of spatially structuring light has enabled a wide range of fundamental studies and applications %\cite{padgett2017orbital}
, including super-resolution microscopy \cite{hell1994breaking}, advanced linear and nonlinear optical interactions with matter \cite{buono2022nonlinear, barros2024observation}, novel light–atom interactions \cite{%tabosa1999optical, 
schmiegelow2016transfer}, high-dimensional quantum information processing \cite{erhard2018twisted}, and ultra–large-bandwidth optical communication \cite{willner2021orbital}, among many others. 
Alongside spatial phase and amplitude engineering, the inclusion of polarization — when the transverse spatial structure is described as a vectorial rather than a scalar field — has further expanded the scope of structured-light research toward spatially varying polarization patterns. 
Vectorially structured light gives rise to rich geometric and topological features such as Möbius strips \cite{bauer2015observation}, polarization knots \cite{larocque2018reconstructing}, and optical skyrmions \cite{shen2024optical}, while also enabling applications in nonlinear and quantum optics \cite{pinheiro2022spin, fickler2014quantum, ornelas2024non}, and stimulating discussion on similarities and differences between classical and quantum nonseparability \cite{spreeuw1998classical,paneru2020entanglement,korolkova2024operational}.\par

The rapid development of the field has been closely linked to technological progress in advanced light-manipulation schemes. 
Tools for spatial light modulation have evolved from holograms \cite{bazhenov1990laser}, phase plates \cite{beijersbergen1994helical}, and computer-controlled spatial light modulators \cite{forbes2016creation} to custom-developed shaping devices based on the geometric phase of light, such as liquid-crystal plates \cite{rubano2019q}, their laser-written counterparts \cite{beresna2011radially}, and nanofabricated metasurfaces \cite{karimi2014generating,dorrah2022tunable}. 
While these techniques have greatly enhanced spatial modulation capabilities, most demonstrations remain limited to single-plane modulation, allowing efficient phase control but providing amplitude control only through loss and preventing the realization of general unitary operations across a set of spatial modes.

To overcome this limitation, schemes implementing multiple phase-modulation planes combined with free-space propagation, so-called multi-plane light conversion (MPLC) architectures, were developed \cite{labroille2014efficient}. 
Propagation between modulation planes enables simultaneous shaping of the phase and amplitude of sets of input and output fields in a theoretically lossless and unitary manner. 
This architecture enables a broad range of operations across large spatial mode spaces with high fidelity, and has rapidly become a major tool for mode-division multiplexing in optical communication \cite{fontaine2019laguerre}, turbulence correction \cite{fontaine2019digital}, spatio-temporal structuring of vectorial light \cite{mounaix2020time}, super-resolved imaging \cite{boucher2020spatial}, advanced quantum photonics applications \cite{brandt2020high, hiekkamaki2021high, lib2022processing, %lib2025experimental,
lib2025high}, and optical neural networks \cite{lin2018all}, among others. 
Because conventional free-space MPLC systems remain bulky and difficult to align \cite{lib2025building}, recent efforts have aimed at miniaturizing MPLC architectures using laser-controlled 3D printing in polymer \cite{wang2025ultracompact}, while being polarization insensitive and, thus, limited to scalar (uniformly polarized) beam manipulation. 
Fully vectorial spatial transformations have also been demonstrated using cascaded reflective metasurfaces \cite{soma2025complete} relying on complex nanofabrication and careful alignment.\par

In this article, we present a compact volumetric multi-plane light conversion scheme realized in standard fused silica glass, capable of jointly modulating the phase, amplitude, and polarization of light within an efficient monolithic architecture. 
The planes are implemented by femtosecond direct laser writing of birefringent nanogratings with a spatially varying orientation of the fast-axis. These modulation planes are <1~mm in transverse dimensions and distributed over a few tens of millimeter propagation distances within the substrate. We first demonstrate high-dimensional spatial unitary transformations at 808 nm wavelength through the implementation of high-dimensional quantum logic, astigmatic mode converter, multimode beam-splitting operation, and mode-independent wave-guiding, thus establishing a compact toolbox for integrated multipath–multimode optical networks. 
We then realize polarization-controlled spatial transformations and the manipulation of complex polarization topologies such as optical skyrmions. 
Finally, we demonstrate the promise of the approach for telecom applications through a 15-mode spatial mode sorter as well as 12-mode polarization and spatial mode sorter. 

Hence, our scheme provides a scalable route towards advanced integrated optical processors for communication, imaging, sensing, and high-dimensional photonic information processing in classical and quantum applications.\\

\begin{figure*}
    \centering
    \includegraphics[width=1\linewidth]{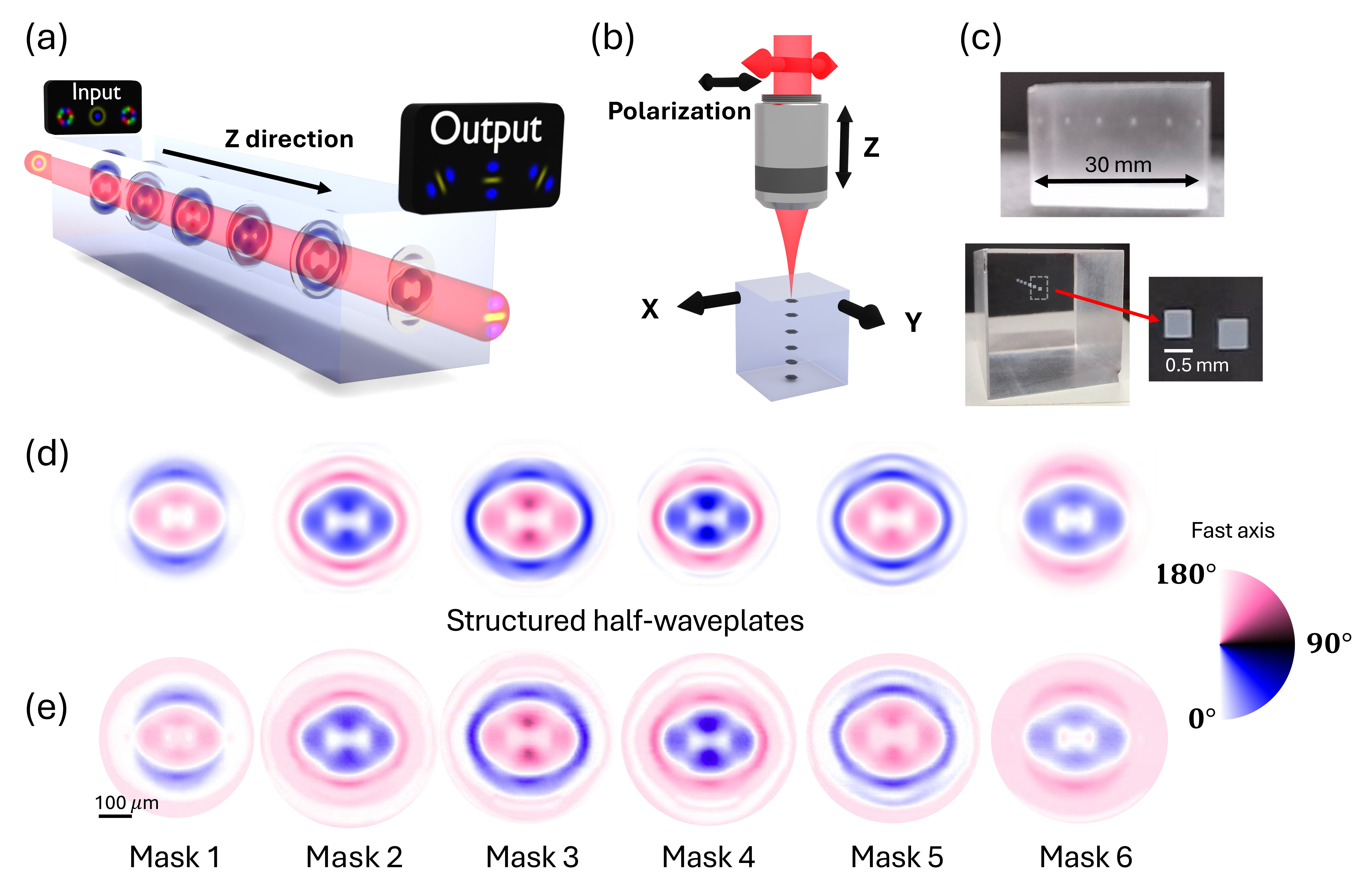}
    \caption{\textbf{Concept and fabrication of volumetric geometric-phase MPLC in glass.} (a) Concept of a volumetric MPLC implemented in a monolithic glass chip, illustrated for a 3D Hadamard gate. (b) Schematic of the femtosecond laser-writing process: the fused-silica sample is translated in $x$--$y$, the objective in $z$, while the input linear polarization is rotated. (c) Photos of a fabricated device showing multiple written planes inside the glass from a side view (top), front view (bottom), and a zoom of 2 planes (inset). (d) Designed fast-axis patterns of the half-waveplate masks for the 3D Hadamard gate. (e) Reconstructed fast-axis distributions of $n$ fabricated masks at depths $d=(n-1)\cdot 5.6$ mm.}
    \label{Fig_1}
\end{figure*}

% \noindent \textbf{Results}\\
\section*{\textbf{Results}}
Femtosecond direct laser writing has been widely used for light-field control in transparent materials \cite{beresna2014ultrafast}, and has already established a versatile platform for applications such as optical data storage \cite{microsoft2026laser} and waveguide-based optical processing \cite{zhang2025recent}. 
The approach presented here exploits the modification regime resulting in polarization-controlled sub-wavelength structures in silica glass.
With this method, ultrashort laser pulses inscribe self-assembled sub-wavelength lamella structures at their focal spot, inducing form birefringence in the glass substrate \cite{shimotsuma2003self}. 
Depending on the laser-writing parameters, different types of birefringence-inducing modifications can be generated, including periodic nanogratings (Type II modification) and low-scattering anisotropic nanoporous structures (Type $X$ modification) \cite{sakakura2020ultralow}. 
In the present work, we employ Type II nanograting modifications because they enable $\pi$ retardance within only 2-3 layers, allowing each modulation plane to operate effectively as a spatially structured half-waveplate. 
By locally varying the fast-axis orientation of these half-waveplates across the transverse plane, the transmitted field acquires a geometric (Pancharatnam–Berry) phase \cite{jisha2021geometric}. 
As a consequence, each modulation plane simultaneously reverses the handedness of circular polarization and imprints opposite conjugate phase profiles to each polarization component, enabling full vectorial control of the spatial field. %  Additional fabrication details are provided in Methods.
%Achieving comparable half-waveplates with Type $X$ modifications would require stacking tens of layers, which increases fabrication complexity and leads to cumulative errors across the volumetric structure.
Additional fabrication details are provided in Methods.

In the present implementation, we achieve transmission efficiencies, i.e. the ratio between the transmitted and incident power of the light, of approximately 89\% per modulation plane at 808 nm and approximately 94\% at 1550 nm. 
%We note that higher efficiencies approaching 99\% have been reported for alternative Type $X$ modification regimes \cite{sakakura2020ultralow}. 

\subsection*{Laser-written MPLC in silica glass}
So far, laser-written geometric-phase elements have relied on single-plane modulation or cascading of a few closely-spaced modulation planes to improve the quality of the generation of a particular mode \cite{chavilkkadan2025purely}.
In contrast, we extend this concept to multiple consecutive modulation planes in combination with some propagation between the planes inside a bulk silica substrate, resulting in the ability to obtain a compact volumetric MPLC architecture as shown in Fig. Fig.~\ref{Fig_1}(a).

The laser-written MPLC examples presented in this article consist of up to 10 modulation planes inscribed within a volume of approximately $0.7 $mm$ \times 0.7 $mm$ \times (10$–$30)$mm, depending on the required inter-plane spacing and device geometry. 
Each modulation plane contains between $200 \times 200$ and $500 \times 500$ pixels, depending on the specific transformation implemented. 
The pixel sizes range from approximately $2\,\mu$m to $8\,\mu$m (exact details of all modulations are given in the supplementary material). 
During the writing of one plane, the glass was raster-scanned through the writing beam with its polarization synchronously modulated to inscribe the half-wave retardance with the desired, spatially-varying fast-axis.  
After writing one modulation plane, the focus of the laser was translated to successive depths inside the glass by moving the objective along the writing-beam direction until all planes were inscribed. 
See Fig.~\ref{Fig_1}(b) for a conceptual image and Fig.~\ref{Fig_1}(c) for exemplary photos of a 6-plane MPLC inscribed in glass.
To maintain the required retardance across the device thickness, two depth-writing strategies were used: first, a depth-dependent pulse-energy compensation (pixel size up to $\sim 8,\mu$m), and second, a slightly divergent beam enabling an approximately constant pixel size of $\sim 2,\mu$m across the full writing depth (see Methods).
A comparison between the ideal structured half-waveplate masks (Fig.~\ref{Fig_1}(d)) and the experimentally reconstructed fast-axis distributions (Fig.~\ref{Fig_1}(e)), written at different depths in glass using the second depth-writing strategy, demonstrates a very good agreement between the designed and fabricated structured half-waveplates. 
The optimization algorithm used to obtain the required transverse polarization modulation is based on wavefront matching \cite{hashimoto2005optical, fontaine2019laguerre}, which we extended to include the vectorial nature of light while also accounting for the constraints imposed by geometric-phase modulation (see Methods). 
For each MPLC implementation, the number of planes, inter-plane distances, and beam size are optimized numerically to achieve the best trade-off between transformation fidelity, modal crosstalk, and a visual inspection of the smoothness of the modulation patterns. 
The latter provides an indication of the robustness of the experimentally implemented transformations with respect to fabrication imperfections.
To test and characterize the performance of the MPLC devices, we generate spatially-structured input fields using i) holographic phase and amplitude modulation with a spatial light modulator for scalar beams \cite{forbes2016creation} and ii) an interferometric scheme for spatial vector beam generation \cite{fickler2014quantum}.
The resulting structured light fields are relayed into the glass substrate using an imaging system. 
The complex optical fields at the output of the MPLC are reconstructed using i) off-axis holography with a reference beam \cite{verrier2011off} for scalar fields, and ii) spatially-resolved polarization tomography for transformations involving polarization patterns (see Supplementary Information).
%The quality of the implemented transformations is evaluated using several metrics formally defined in the Methods section. 
The reconstructed optical fields are qualitatively compared with the corresponding ideal target modes to verify the expected spatial structures. 
As a quantitative measure of the transformation fidelity and modal selectivity, we evaluated the visibility of reconstructed crosstalk matrices obtained from overlap measurements with the ideal modes (see Methods). 

\begin{figure}
    \centering
    \includegraphics[width=1\linewidth]{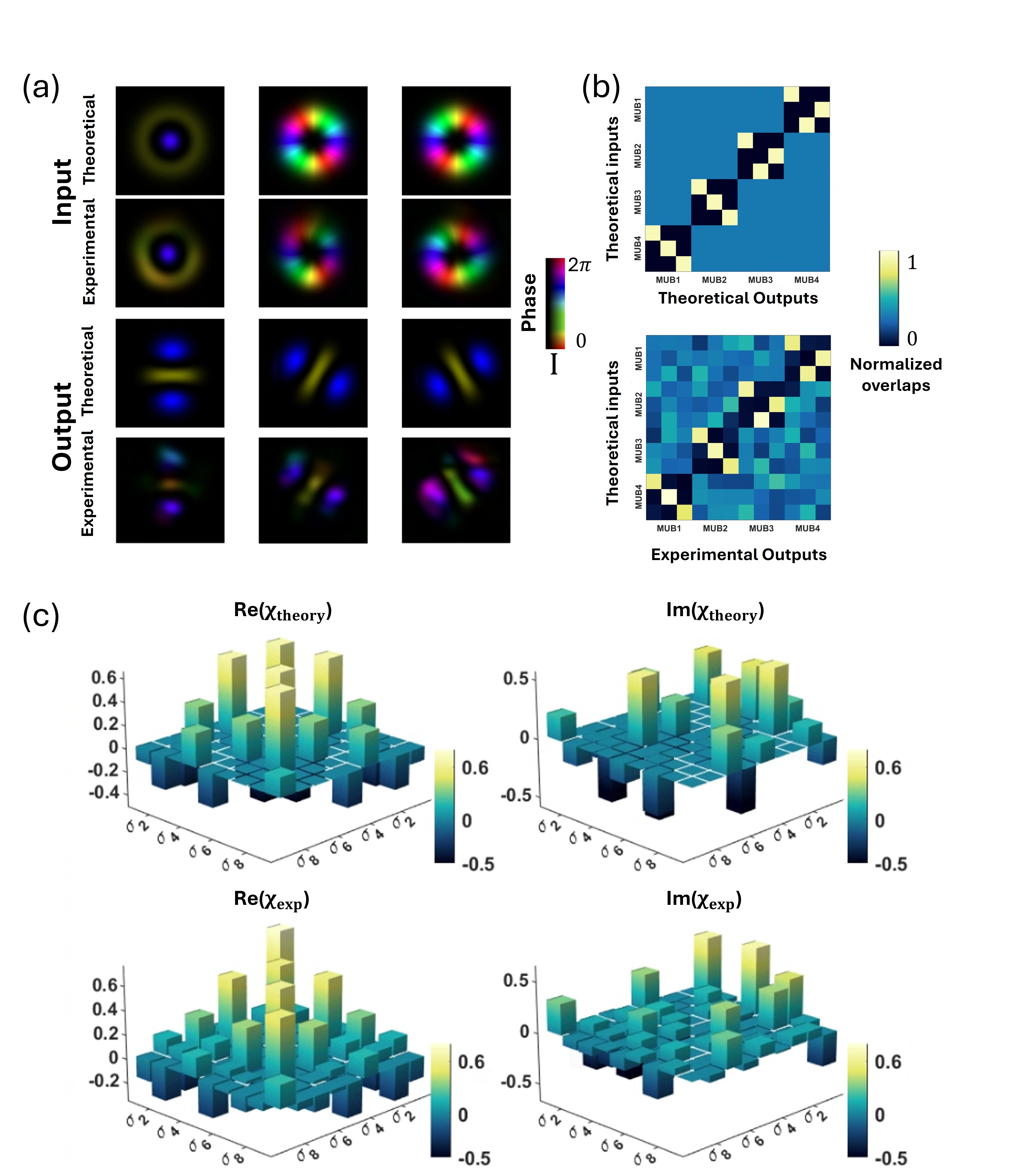}
    \caption{\textbf{Laser-written 3D Hadamard quantum gate in glass.} (a) Theoretical input and output modes for the 3D Hadamard gate together with their experimentally reconstructed counterparts. (b) Ideal (top) and measured (bottom) normalized crosstalk matrices obtained by probing all mutually unbiased basis (MUB) states. (c) Real and imaginary parts of the theoretical and measured process $\chi$ matrix of the 3D Hadamard gate, reconstructed using the Gell–Mann matrix basis.}
    \label{Fig_2}
\end{figure}

\subsection*{Laser-written MPLC for scalar light field modulations}
In a first set of experiments, we show the modulation ability of the laser-written MPLC platform for scalar light fields at 808 nm, using a few example transformations in promising applications.

\begin{figure*}[ht]
    \centering
    \includegraphics[width=1\linewidth]{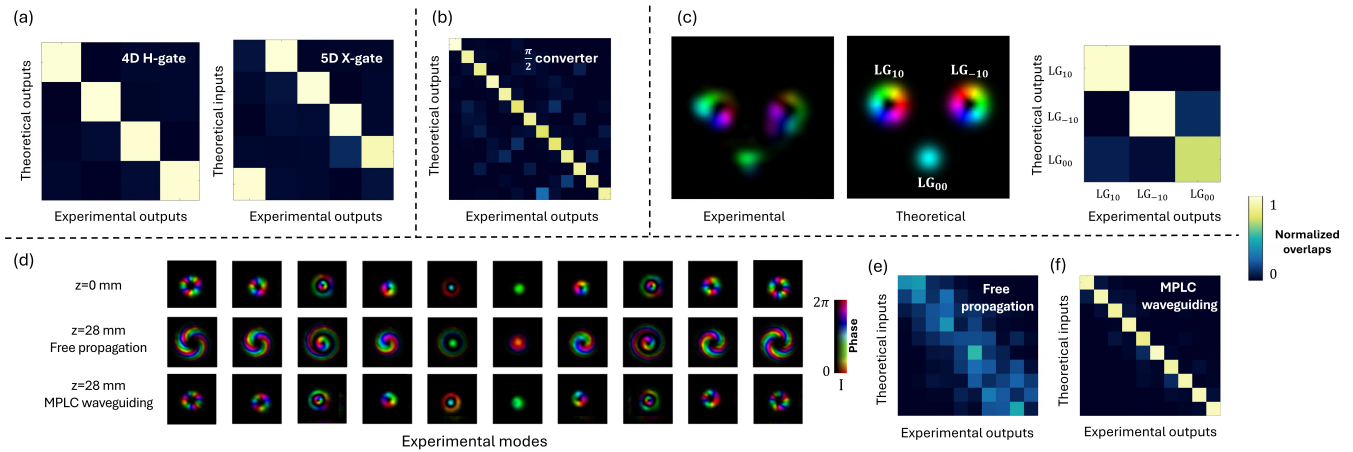}
    \caption{\textbf{Scalar spatial transformations implemented with volumetric MPLC.} (a) Normalized crosstalk matrix for a 4D Hadamard gate and unitary matrix for a 5D $X$-gate constructed from overlap values. (b) Crosstalk matrix for the astigmatic $\pi/2$ mode conversion from the 13 lowest-order LG to HG modes. (c) Experimentally reconstructed output fields of the multimode beam splitter for an input Gaussian beam together with the corresponding normalized crosstalk matrix. (d) Experimentally measured LG input modes at the entrance of the MPLC waveguiding device ($z=0$), their freely propagated counterparts after $z=28$\,mm in glass, and the corresponding output modes when guided through the glass by the MPLC. Normalized crosstalk matrix without (e) and with (f) MPLC waveguiding.}
    \label{Fig_3}
\end{figure*}

\textbf{High-dimensional spatial unitary transformations}:
To demonstrate high-dimensional spatial unitary transformations, we implemented single-beam-line 3D and 4D Hadamard-gates, as well as a 5D cyclic permutation (X-gate). 
For the 3D Hadamard transformation, the Laguerre–Gaussian modes  LG$_{2,0}$, LG$_{-2,0}$, and LG$_{0,1}$ were used, where the indices of LG$_{\ell,p}$ label the azimuthal ($\ell$) and radial ($p$) index.  By probing all states of all four mutually unbiased bases, a full quantum process tomography was performed \cite{bouchard2019quantum}.
The experimentally reconstructed output modes are in very good agreement with the simulated results (Fig.~\ref{Fig_2}(a)).
The corresponding normalized crosstalk matrix for all mutually unbiased basis (MUB) states is shown in Fig.~\ref{Fig_2}(b) with a visibility of 
92\% and a very good resemblance to the one expected from theory. 
The real and imaginary part of the reconstructed process matrices, along with the theoretical ones, can be seen in Fig.~\ref{Fig_2}(c). 
From these measurements, we obtained a process fidelity of 83\% and a process purity of 89\%.
For the 4D Hadamard-gate and 5D X-gate, we used the LG modes sets of \{LG$_{3,0}$, LG$_{-3,0}$, LG$_{1,1}$, LG$_{-1,1}$\} and \{LG$_{-2,0}$, LG$_{-1,0}$, LG$_{0,0}$, LG$_{1,0}$, LG$_{2,0}$\}, respectively.
The measured crosstalk matrix (4D Hadamard-gate) and reconstructed unitary transformation matrix (5D X-gate) are shown in Fig.~\ref{Fig_3}(a).
In both cases, the visibilities of the respective matrices exceeds 90\%.

\textbf{Astigmatic mode conversion}:
For the spatial mode transformation between LG modes and Hermite-Gaussian modes, a bulk optics solution consisting of a cylindrical lens pair is known \cite{beijersbergen1993astigmatic}.
We show that the same functionality can achieved in our intrinsically aligned, laser-written MPLC plattform.
Specifically, we implemented the mode conversion transforming a set of 13 LG modes into the corresponding HG modes using 2 modulation planes. 
The measured crosstalk matrix exhibits low modal crosstalk with a visibility of 90\%, confirming high-fidelity mode conversion (see Fig.~\ref{Fig_3}(b) and supplementary for more details).

\textbf{Multimode beam splitting}:
We further demonstrate multimode beam-splitting operations, where a single input Gaussian mode is coherently split into three output paths each having a different spatial mode, i.e., LG$_{1,0}$, LG$_{-1,0}$, and LG$_{0,0}$. 
The corresponding normalized crosstalk matrix between the three output channels is shown in Fig.~\ref{Fig_3}(c), yielding an average visibility of 87\%.
The combination of beam splitting and simultaneous mode transformation might be interesting for more complex multi-path and multi-mode network structures integrated into a single silica glass.

\textbf{LG mode-independent waveguiding via geometric phases}:
The nanograting-based MPLC naturally matches an unconventional approach to waveguiding of light based on geometric phase control instead of refractive-index modulations, which was recently demonstrated for bulk optics devices \cite{slussarenko2016guiding}. 
We extend this concept to mode-independent guiding of light implemented through 10 modulation planes through 28 mm of glass.
Fig.~\ref{Fig_3}(d) shows ten different experimentally reconstructed input modes at the beginning of the device (z=0) in comparison with the same modes after free propagation through glass and their guided counterparts. 
Their respective normalized crosstalk matrices nicely demonstrate the geometric-phase guiding (see Fig.~\ref{Fig_3}(e) and Fig.~\ref{Fig_3}(f)).
More quantitatively, the matrix of the guided modes yields an average visibility exceeding 90\%, while the one for unguided modes drops to 26\%.

Although the presented results for advanced scalar beam modulations only present a very small subset of all possible multi-mode transformations, these examples highlight the vast potential of an integrated MPLC as a toolbox for larger multimode-multipath optical networks.

%%%%%%%%%%%%%%%%%%%%%%%
\subsection*{Laser-written MPLC for vectorial light field modulations}
In addition to leveraging the laser-written MPLC as a scalar spatial processing platform, we also exploited its ability to modulate polarization to realize full vectorial light field transformations.

\begin{figure}
    \centering
    \includegraphics[width=1\linewidth]{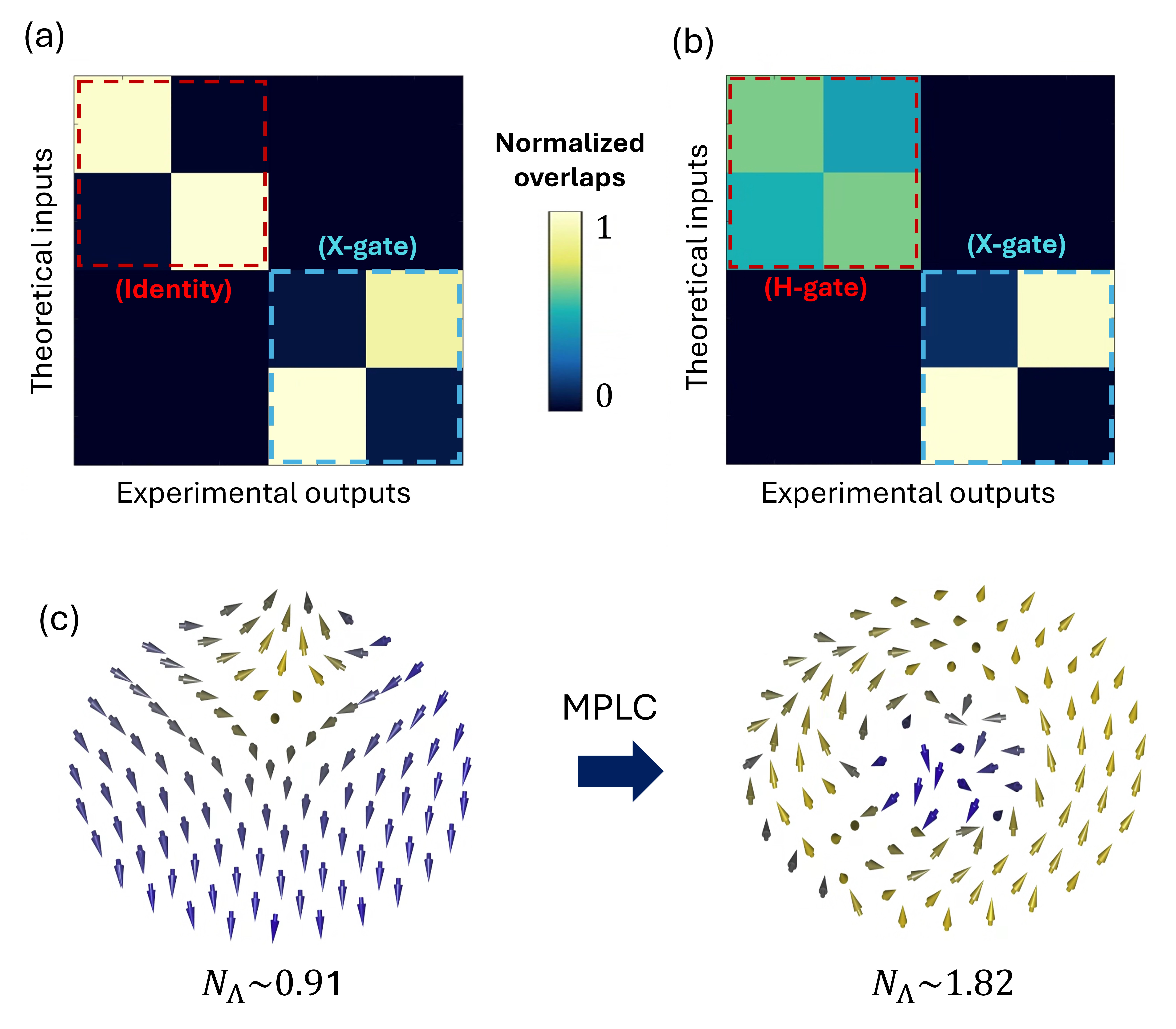}
    \caption{\textbf{Polarization-controlled spatial unitary transformations and skyrmion topology conversion.} (a) Measured unitary matrix of the polarization-controlled cNOT gate. (b) Measured unitary matrix showing polarization-controlled switching between a Hadamard-gate ($H$) and a NOT-gate ($X$). (c) Reconstructed Stokes-vector distributions of the optical skyrmion before and after the transformation by the MPLC and their respective skyrmion numbers.}
    \label{Fig_4}
\end{figure}

\textbf{Polarization-controlled spatial unitary transformations}:
At first, we implemented a two-qubit controlled NOT (cNOT) operation in which the polarization state acts as the control qubit and the orbital angular momentum (OAM) represents the target qubit. 
In this case, right-circularly polarized light undergoes a NOT operation (X-gate) between the two OAM modes, whereas left-circularly polarized light experiences the identity operation. 
The transformation was demonstrated for the input mode set LG$_{1,0}^{R}$, LG$_{-1,0}^{L}$, LG$_{1,0}^{L}$, and LG$_{-1,0}^{R}$, with the corresponding measured unitary transformation matrix shown in Fig.~\ref{Fig_4}(a).
Beyond the realization of a cNOT gate, the same scheme can also be used to leverage polarization for multiplexing different unitary operations within a single beam path. As an example, we implemented a device in which a two-dimensional Hadamard ($H$) transformation is applied for left-circular polarization, while a two-dimensional cyclic permutation ($X$) transformation is applied for right-circular polarization using the same input mode set as in the previous demonstration. The corresponding measured unitary transformation matrix are shown in Fig.~\ref{Fig_4}(b). 
For both polarization-controlled transformations, low intermodal crosstalk is observed, with visibilities exceeding 94\%.

\textbf{Transformation of optical skyrmions}:
The ability to fully control the polarization across the transverse spatial plane, further enabled us to extend the use of the laser-written MPLC platform to transform light with complex topologies such as optical skyrmions.
Optical skyrmions are formed through non-separable states between polarization and spatial modes of different transverse amplitude profiles.
The topology of the resulting spatially structured polarization textures can be characterized by the skyrmion number \cite{shen2024optical} and has attracted a lot of attention for its potential use as a resilient carrier of classical \cite{zhang2025topological} and quantum information \cite{ornelas2024non, ornelas2025topological}. 
Although optical skyrmions have been extensively studied in the context of their generation and characterization, only a few works have addressed controlled transformations between skyrmion states \cite{liu2025control}. Here, we illustrate this capability by demonstrating the conversion of a light field with skyrmion number 1 into a light field with skyrmion number 2.
The experimentally reconstructed Stokes-vector distributions of the input and transformed skyrmion states are shown in Fig.~\ref{Fig_4}(c). 
The quality of the transformation is evaluated by calculating the corresponding Skyrmion number before and after the transformation, for which we obtain values of 0.93 and 1.83, respectively.
While both values are in good agreement with the theory, experimental imperfection at the generation as well as the transformation result in small discrepancies to the ideal values.

Together with polarization-controlled gate operations, these results demonstrate the capability of the volumetric MPLC architecture to perform a broad range of vectorial mode transformations impossible to achieve with a single modulation plane.

%%%%%%%%%%%%%%%%%%%%%%%%%%
\subsection*{High-dimensional multiplexing at telecom wavelengths}
\begin{figure}
    \centering
    \includegraphics[width=1\linewidth]{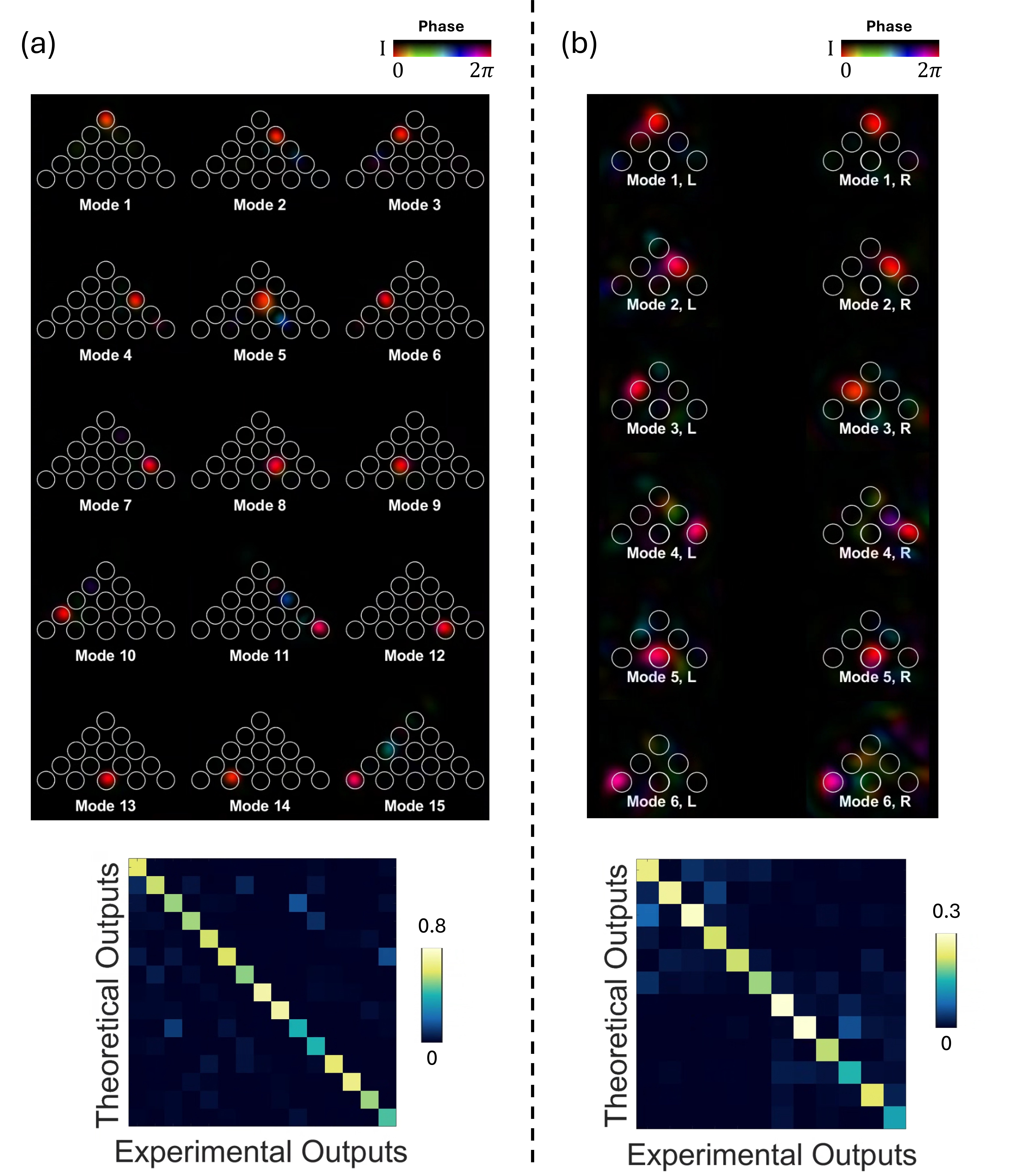}
    \caption{\textbf{Telecom-wavelength Hermite--Gaussian multiplexing.} (a) Experimentally reconstructed output fields of a 15-mode HG sorter, showing spatial separation of all HG modes up to mode order five de-multiplexed into distinct output positions. The corresponding measured (non-normalized) crosstalk matrix is shown below. (b) Experimental reconstruction of the output light fields of a polarization and Hermite–Gaussian (HG) mode sorter for six spatial modes and left (L) and right (R) circular polarization. The corresponding measured crosstalk matrix (non-normalized) is shown below. For both sorters, the logarithmic representation of the crosstalk matrix is provided in the Supplementary}
    \label{Fig_5}
\end{figure}

Finally, to demonstrate wavelength scalability and compatibility with optical communication platforms, we implemented high-dimensional de-multiplexing transformations at the telecom wavelength of 1550~nm.
At first, a 15-dimensional Hermite--Gaussian (HG) de-multiplexer (mode sorter) for scalar light fields was implemented.
Here, every input mode was transformed into a localized Gaussian output at a distinct spatial output position. 
At this wavelength, the laser-written birefringent modulation plane provided a retardance of approximately $0.85\pi$, leading to a moderate reduction in transformation efficiency of around 10\% according to simulations, with no change in visibility compared to the ideal half-wave condition. 
A full $\pi$ retardance can in principle be recovered by adding an additional layer of nanogratings per plane. 
However, this would introduce further accumulated fabrication errors across the device, potentially degrading the overall performance more than the moderate efficiency reduction.
Fig.~\ref{Fig_5}(a) shows the experimentally reconstructed complex optical output fields together with their spatial output locations. 
We calculated the overlap of each output spot with a Gaussian mode (mimicking single-mode fiber coupling) and reconstructed the resulting crosstalk matrix (Fig.~\ref{Fig_5}(a)). An average visibility of 81\% is obtained, with intermodal crosstalk below approximately $-7.9$\,dB. 
The diagonal elements corresponding to the desired input–output mapping exhibit an average Gaussian-mode overlap of approximately 56\%, which equates to an SMF coupling loss of about $-2.6$\,dB.

Similar to before, through including the modulation of polarization, we extended the functionality and implemented a polarization-resolved HG mode sorter with a dimensionality of 12 modes.
In this configuration, six HG modes are routed to the right side of the output grid for right-circular polarization and to the left side for left-circular polarization.
As for the scalar case, we reconstructed the complex output fields for both polarization components together with the corresponding crosstalk matrix (see Fig.~\ref{Fig_5}(b)). 
The measured transformation yields an average visibility of 76\% with intermodal crosstalk below approximately $-11.2$\,dB. 
The diagonal elements correspond to an average Gaussian overlap of approximately 25\%, equivalent to a coupling loss of about $-6.5$\,dB. % relative to ideal single-mode fiber coupling. 
The reduced performance compared to the scalar sorter originates from the combined effect of polarization-dependent routing and the non-ideal birefringent retardance at this wavelength.
While this can be accounted for in the future, the results already show the potential of our scheme for an efficient, compact, and versatile device for optical communications at telecom wavelengths \cite{fontaine2019laguerre} along with the mode decompositions required in super-resolved imaging 
\cite{pushkina2021superresolution}.

%% discussion 

\section*{Discussion}

In this work, we demonstrated a monolithic geometric-phase MPLC architecture in fused-silica based on femtosecond laser-written birefringent nanogratings, enabling compact vectorial light manipulation and high-dimensional spatial transformations.
%We demonstrated its versatility in various applications for scalar as well as vectorial lights in tasks such as high-dimensional quantum gates, unconventional waveguiding, advanced control over optical topologies, and de-multiplexing for optical communications. 
The presented modulations are enabled through volumetric birefringent modulation written at depths of up to 3~cm while maintaining an effective transverse resolution of approximately $2\,\mu$m using depth-compensated writing strategies. 
With the present experimental setup, writing depths up to approximately 4.5~cm are accessible, with further extensions possible at the expense of transverse pixel resolution.
Increasing the accessible writing depth will enable scaling to transformations involving larger sets of modes by allowing larger numbers of modulation planes. 
Unlike bulk optics MPLC systems, requiring careful multi-plane alignment, the addition of more modulation planes does not increase the complexity of its implementation, due to the intrinsic alignment of the modulation planes.

In our current implementations, the main limiting factor in the MPLC's performance are the slight imperfections in the synchronization between the translation stages and electro-optic control of the polarization of the writing laser during fabrication (see Supplementary Information for more details). 
Improved hardware synchronization will result in a more accurate spatial resolution and matching of the written and the desired modulation patterns.
Another performance metric, i.e., the transmission efficiency of each modulation plane, can be improved through slight adjustments of our setup and write the birefringence modulation planes using anisotropic nanoporous structures where values up top 99\% efficiency have been shown \cite{sakakura2020ultralow}. 
In addition, an interesting approach in the future will be to extend the MPLC's modulation capability to inscribe arbitrary retardance values. 
This will provide an additional parameter to optimize, which we expect to not only further improve the modulation performance, but also enable more general transformations.
We further anticipate that spatio-spectral mode and polarization processing within a single substrate will become feasible through including dispersive or spectrally selective elements in the device. 
Such an extension will increase the applicability to high-dimensional classical and quantum information processing and multiplexing also invoking the spectral and temporal degree of freedom of light.
Lastly, since our scheme of using birefringence engineering works for standard glass substrates it can readily be used for example to enhance the performance of glass interfaces through adding custom-tailored spatial mode matching techniques. 

\vspace{5mm}
We note that during the preparation of this manuscript, we became aware of a recent preprint reporting a two-plane geometric-phase MPLC implementation based on laser-written nanogratings which demonstrates a 10-mode scalar spatial mode sorter with discrete phase modulation steps \cite{butaite2026miniaturised}.

\section*{Acknowledgments}
\noindent
We thank Marco Ornigotti and Rafael Barris for the valuable discussions, Pedro Ornelas for support with the code for skyrmion reconstruction, and Alexander Szameit and his group for valuable information on laser writing techniques. RF acknowledges the support of the Research Council of Finland through the Academy Research Fellowship (decision 332399). OK and RF acknowledge the support of the Research Council of Finland through the project BIQOS (decision 336375 and 358134) and the Photonics Research and Innovation Flagship (PREIN - decision 346511). 

\section*{Author contributions}
\noindent
The idea was conceived by RF. The experimental setup was designed by OK and RF. The code for obtaining the required modulations was developed by OK and MH. The setup was built by OK. All measurements and data analysis were performed by OK. The manuscript was written by OK and RF, with edits by MH. RF supervised the study.

\section*{Competing interests}
\noindent
The authors declare no competing interest

\section*{Methods}
\subsection*{Laser writing and fabrication process}
The modulation planes used in this work were fabricated inside fused silica using a femtosecond laser direct writing technique. 
A fiber laser operating at a wavelength of 1030~nm, with a pulse duration of approximately 200~fs, a repetition rate of 250~kHz, and an average power in the range of 150–350~mW, was used as the writing source.  
The laser beam was expanded to a diameter of approximately 1~cm to efficiently fill the entrance aperture of a long-working-distance microscope objective with a numerical aperture of 0.20, which focused the beam inside the glass substrate. 
The objective was mounted on a motorized translation stage along the optical axis ($z$-axis in Fig.~\ref{Fig_1}), allowing precise positioning of the focal spot at different depths inside the glass. The fused silica sample was mounted on high-precision motorized translation stages enabling transverse ($xy$) positioning during the writing process.
The laser focus was translated through the substrate to inscribe arrays of laser-induced nanogratings either by moving the glass ($xy$ plane) or by moving the objective ($z$ direction). 
The nanogratings induce birefringent structures, whose local fast-axis orientation is perpendicular to the orientation of the linear polarization direction of the writing beam. 
By varying the laser's polarization during fabrication, a birefringent layer with a designed fast-axis distribution can be produced \cite{beresna2014ultrafast}. 
In the present work, each modulation plane was realized as a continuous distribution of nanogratings acting as a spatially varying half-waveplate with an induced retardance of $\pi$.
The modulation planes were written at depths of up to 3~cm inside the fused silica substrate, starting from the deepest layer and proceeding toward the surface.
Each modulation plane was implemented as two stacked nanograting layers separated by 100~$\mu$m in order to achieve a total retardance of approximately $\pi$ at 808~nm and $0.87\pi$ at 1550~nm. 
The nanograting structures were inscribed by raster scanning through the glass substrate across the transverse plane relative to the focused beam. 
For each layer, 500 or 700 parallel lines were written with a line spacing of 1~$\mu$m and a line length of 0.5 or 0.7~mm, at a translation speed of 0.2~mm/s.
A detailed description of the laser-writing setup, the obtained retardance, and how we experimentally characterized it is provided in the Supplementary Information.

\subsection*{Depth-dependent calibration of femtosecond laser writing parameter}
Two different fabrication strategies were used to compensate for depth-dependent focal spot sizes.
In the first strategy, the writing beam was collimated before entering the microscope objective. 
When focusing deeper into the sample, the refractive index of glass enlarges the focal distance thereby reducing the focal spot size and local intensity, leading to decreased birefringent retardance. 
To compensate for this effect, the pulse energy was progressively increased with writing depth. 
While this approach preserves the required retardance, it also results in a depth-dependent transverse pixel size. 
In the second strategy, the writing beam was made slightly divergent before entering the microscope objective. As the objective position was translated from $z_0$ to deeper positions inside the glass, the beam diameter at the entrance aperture increased, compensating for focal broadening caused by the refractive index of glass. As a result, a nearly constant focal-spot size and retardance close to $\pi$ were maintained over the full writing depth. 
Further characterization of these depth-compensation strategies is provided in the Supplementary Information.

%%%%

\subsection*{Wavefront-matching optimization for structured half-waveplates}
The structured half-waveplates at each modulation plane were optimized using a vectorial extension of the wavefront-matching algorithm \cite{fontaine2019laguerre}. In contrast to the conventional scalar formulation, where modulation planes introduce spatial phase delays, the present approach determines the spatial distribution of the fast-axis orientation $\theta_t(x,y)$, which controls the geometric phase applied to the optical field at $t$ modulation planes along the propagation direction.

The forward- and backward-propagated optical fields are expressed as scalar spatial modes multiplied by polarization vectors. For each input mode $r$ and target mode $s$, the fields can be written as
\begin{equation}
\mathbf{F}_r(x,y) = f_r(x,y)\,\mathbf e_\sigma,
\qquad
\mathbf{B}_s(x,y) = b_s(x,y)\,\mathbf e_\sigma,
\end{equation}
where $f_r(x,y)$ and $b_s(x,y)$ denote the scalar spatial mode distributions and $\mathbf e_\sigma$ represents the circular polarization state.

For generality, the fields are expressed in the linear $H$–$V$ polarization basis as Jones vectors
\begin{equation}
\mathbf{F}_r(x,y)=
\begin{pmatrix}
F_{r,H}(x,y)\\
F_{r,V}(x,y)
\end{pmatrix},
\qquad
\mathbf{B}_s(x,y)=
\begin{pmatrix}
B_{s,H}(x,y)\\
B_{s,V}(x,y)
\end{pmatrix},
\end{equation}
which are used for evaluating modal overlaps during the optimization procedure.

For circular polarization, the fields acquire the geometric-phase modulation introduced by structured half-waveplates. A half-wave plate with local fast-axis orientation $\theta_t(x,y)$ flips the handedness of the circular polarization and introduces a geometric phase $\Phi_t(x,y)=\pm 2\theta_t(x,y)$.

During optimization, the input modes are propagated forward and the target modes backward through the system using split-step propagation in glass, while modulation masks are applied at their corresponding plane positions. At each modulation plane $t$, the geometric-phase update is obtained from the overlap between forward- and backward-propagated fields projected onto the circular polarization components,
\begin{equation}
o_R=\frac{1}{2}(B_{s,H}^*(x,y)+ iB_{s,V}^*(x,y))(F_{r,H}(x,y)+iF_{r,V}(x,y)),
\end{equation}
\begin{equation}
o_L=\frac{1}{2}(B_{s,H}^*(x,y)-iB_{s,V}^*(x,y))(F_{r,H}(x,y)-iF_{r,V}(x,y)).
\end{equation}

After removing global phase offsets,
\begin{equation}
\phi_R=\arg(\langle o_R\rangle),
\qquad
\phi_L=\arg(\langle o_L\rangle),
\end{equation}
the update term is written as
\begin{equation}
o_{rs}=o_R e^{-i\phi_R}+o_L^* e^{i\phi_L}.
\end{equation}

The geometric-phase correction at plane $t$ is then obtained as
\begin{equation}
\Delta\Phi_t(x,y)=-\arg\left(\sum_{r,s} o_{rs}(x,y,t)\right),
\end{equation}
from which the fast-axis orientation follows as
\begin{equation}
\theta_t(x,y)=\tfrac{1}{2}\Delta\Phi_t(x,y)\ (\mathrm{mod}\ \pi).
\end{equation}

The retrieved fast-axis orientation at each iteration is added to the previously accumulated distribution of the modulation plane until convergence of the mode overlaps is achieved.

\subsection*{Performance metrics}
To quantify the performance of the implemented transformations, modal overlaps between experimentally reconstructed output fields and the corresponding ideal target modes were calculated as
\begin{equation}
C_{ij} =
\frac{\left|\langle E_i^{\mathrm{exp}} \mid E_j^{\mathrm{out}} \rangle \right|^2}
{\langle E_i^{\mathrm{exp}} \mid E_i^{\mathrm{exp}} \rangle
\langle E_j^{\mathrm{out}} \mid E_j^{\mathrm{out}} \rangle},
\end{equation}

where $E_i^{\mathrm{exp}}$ denotes the experimentally reconstructed output field corresponding to input mode $i$, and $E_j^{\mathrm{out}}$ represents the ideal target mode $j$. 
The coefficients $C_{ij}$ define the elements of the crosstalk matrix.
To evaluate the transformation fidelity independently of losses outside the measured modal subspace, the crosstalk matrix was normalized as
\begin{equation}
\tilde{C}_{ij} =
\frac{C_{ij}}{\sum_{k=1}^{N} C_{ik}},
\end{equation}
so that each row sums to unity.

The visibility of the transformation was calculated from the non-normalized crosstalk matrix as
\begin{equation}
V =
\frac{1}{N}
\sum_{i=1}^{N}
\frac{C_{ii}}{\sum_{j=1}^{N} C_{ij}},
\end{equation}
where $N$ is the number of modes.

In addition, the unitary transformation matrix $U_{ij}$ was reconstructed from the complex modal overlaps between experimentally measured output modes and the corresponding ideal input modes as
\begin{equation}
U_{ij} =
\frac{\left|\langle E_i^{\mathrm{exp}} \mid E_j^{\mathrm{input}} \rangle \right|^2}
{\langle E_i^{\mathrm{exp}} \mid E_i^{\mathrm{exp}} \rangle
\langle E_j^{\mathrm{input}} \mid E_j^{\mathrm{input}} \rangle},
\end{equation}

The reconstructed matrix was normalized using the same procedure as before.

\bibliography{references}

@article{forbes2021structured,
  title={Structured light},
  author={Forbes, Andrew and De Oliveira, Michael and Dennis, Mark R},
  journal={Nature photonics},
  volume={15},
  number={4},
  pages={253--262},
  year={2021},
  publisher={Nature Publishing Group UK London},
  url={https://doi.org/10.1038/s41566-021-00780-4}
  
}

@article{shen2024optical,
  title={Optical skyrmions and other topological quasiparticles of light},
  author={Shen, Yijie and Zhang, Qiang and Shi, Peng and Du, Luping and Yuan, Xiaocong and Zayats, Anatoly V},
  journal={Nature Photonics},
  volume={18},
  number={1},
  pages={15--25},
  year={2024},
  publisher={Nature Publishing Group UK London},
  url={https://doi.org/10.1038/s41566-023-01325-7}
}

@article{he2022towards,
  title={Towards higher-dimensional structured light},
  author={He, Chao and Shen, Yijie and Forbes, Andrew},
  journal={Light: Science \& Applications},
  volume={11},
  number={1},
  pages={205},
  year={2022},
  publisher={Nature Publishing Group UK London},
  url={https://doi.org/10.1038/s41377-022-00897-3}
}

@article{dorrah2022tunable,
  title={Tunable structured light with flat optics},
  author={Dorrah, Ahmed H and Capasso, Federico},
  journal={Science},
  volume={376},
  number={6591},
  pages={eabi6860},
  year={2022},
  publisher={American Association for the Advancement of Science},
  url={https://doi.org/10.1016/j.optcom.2025.132599}
}

@article{hell1994breaking,
  title={Breaking the diffraction resolution limit by stimulated emission: stimulated-emission-depletion fluorescence microscopy},
  author={Hell, Stefan W and Wichmann, Jan},
  journal={Optics letters},
  volume={19},
  number={11},
  pages={780--782},
  year={1994},
  publisher={Optical Society of America},
  url={https://doi.org/10.1364/OL.19.000780}
}

@article{pinheiro2022spin,
  title={Spin to orbital angular momentum transfer in frequency up-conversion},
  author={Pinheiro da Silva, Braian and Buono, Wagner T and Pereira, Leonardo J and Tasca, Daniel S and Dechoum, Kaled and Khoury, Antonio Z},
  journal={Nanophotonics},
  volume={11},
  number={4},
  pages={771--778},
  year={2022},
  publisher={De Gruyter},
  url={https://doi.org/10.1515/nanoph-2021-0493}
}

@article{barros2024observation,
  title={Observation of the topological aberrations of twisted light},
  author={Barros, Rafael F and Bej, Subhajit and Hiekkam{\"a}ki, Markus and Ornigotti, Marco and Fickler, Robert},
  journal={Nature Communications},
  volume={15},
  number={1},
  pages={8162},
  year={2024},
  publisher={Nature Publishing Group UK London},
  url={https://doi.org/10.1038/s41467-024-52529-6}
}

@article{schmiegelow2016transfer,
  title={Transfer of optical orbital angular momentum to a bound electron},
  author={Schmiegelow, Christian T and Schulz, Jonas and Kaufmann, Henning and Ruster, Thomas and Poschinger, Ulrich G and Schmidt-Kaler, Ferdinand},
  journal={Nature communications},
  volume={7},
  number={1},
  pages={12998},
  year={2016},
  publisher={Nature Publishing Group UK London},
  url={https://doi.org/10.1038/ncomms12998}
}

@article{erhard2018twisted,
  title={Twisted photons: new quantum perspectives in high dimensions},
  author={Erhard, Manuel and Fickler, Robert and Krenn, Mario and Zeilinger, Anton},
  journal={Light: Science \& Applications},
  volume={7},
  number={3},
  pages={17146--17146},
  year={2018},
  publisher={Nature Publishing Group},
  url={https://doi.org/10.1038/lsa.2017.146}
}

@article{buono2022nonlinear,
  title={Nonlinear optics with structured light},
  author={Buono, Wagner Tavares and Forbes, Andrew},
  journal={Opto-Electronic Advances},
  volume={5},
  number={6},
  pages={210174--1},
  year={2022},
  url={https://doi.org/10.29026/oea.2022.210174}
}

@article{willner2021orbital,
  title={Orbital angular momentum of light for communications},
  author={Willner, Alan E and Pang, Kai and Song, Hao and Zou, Kaiheng and Zhou, Huibin},
  journal={Applied Physics Reviews},
  volume={8},
  number={4},
  year={2021},
  publisher={AIP Publishing},
  url={https://doi.org/10.1063/5.0054885}
}

@article{bauer2015observation,
  title={Observation of optical polarization M{\"o}bius strips},
  author={Bauer, Thomas and Banzer, Peter and Karimi, Ebrahim and Orlov, Sergej and Rubano, Andrea and Marrucci, Lorenzo and Santamato, Enrico and Boyd, Robert W and Leuchs, Gerd},
  journal={Science},
  volume={347},
  number={6225},
  pages={964--966},
  year={2015},
  publisher={American Association for the Advancement of Science},
  url={https://doi.org/10.1126/science.1260635}
}

@article{larocque2018reconstructing,
  title={Reconstructing the topology of optical polarization knots},
  author={Larocque, Hugo and Sugic, Danica and Mortimer, Dominic and Taylor, Alexander J and Fickler, Robert and Boyd, Robert W and Dennis, Mark R and Karimi, Ebrahim},
  journal={Nature Physics},
  volume={14},
  number={11},
  pages={1079--1082},
  year={2018},
  publisher={Nature Publishing Group UK London},
  url={https://doi.org/10.1038/s41567-018-0229-2}
}

@article{ornelas2024non,
  title={Non-local skyrmions as topologically resilient quantum entangled states of light},
  author={Ornelas, Pedro and Nape, Isaac and de Mello Koch, Robert and Forbes, Andrew},
  journal={Nature Photonics},
  volume={18},
  number={3},
  pages={258--266},
  year={2024},
  publisher={Nature Publishing Group UK London},
  url={https://doi.org/10.1038/s41566-023-01360-4}
}

@article{fickler2014quantum,
  title={Quantum entanglement of complex photon polarization patterns in vector beams},
  author={Fickler, Robert and Lapkiewicz, Radek and Ramelow, Sven and Zeilinger, Anton},
  journal={Physical Review A},
  volume={89},
  number={6},
  pages={060301},
  year={2014},
  publisher={APS},
  url={https://doi.org/10.1103/PhysRevA.89.060301}
}

@article{spreeuw1998classical,
  title={A classical analogy of entanglement},
  author={Spreeuw, Robert JC},
  journal={Foundations of physics},
  volume={28},
  number={3},
  pages={361--374},
  year={1998},
  publisher={Springer},
  url={https://doi.org/10.1023/A:1018703709245}
}

@article{paneru2020entanglement,
  title={Entanglement: quantum or classical?},
  author={Paneru, Dilip and Cohen, Eliahu and Fickler, Robert and Boyd, Robert W and Karimi, Ebrahim},
  journal={Reports on Progress in Physics},
  volume={83},
  number={6},
  pages={064001},
  year={2020},
  publisher={IOP Publishing},
  url={https://doi.org/10.1088/1361-6633/ab85b9}
}

@article{korolkova2024operational,
  title={An operational distinction between quantum entanglement and classical non-separability},
  author={Korolkova, Natalia and S{\'a}nchez-Soto, Luis and Leuchs, Gerd},
  journal={Philosophical Transactions of the Royal Society A: Mathematical, Physical and Engineering Sciences},
  volume={382},
  number={2287},
  year={2024},
  publisher={The Royal Society},
  url={https://doi.org/10.1098/rsta.2023.0342}
}

@article{beijersbergen1994helical,
  title={Helical-wavefront laser beams produced with a spiral phaseplate},
  author={Beijersbergen, MW and Coerwinkel, RPC and Kristensen, M and Woerdman, JP},
  journal={Optics communications},
  volume={112},
  number={5-6},
  pages={321--327},
  year={1994},
  publisher={Elsevier},
  url={https://doi.org/10.1016/0030-4018(94)90638-6}
}

@article{bazhenov1990laser,
  title={Laser beams with screw dislocations in their wavefronts},
  author={Bazhenov, V Yu and Vasnetsov, MV and Soskin, MS},
  journal={Optical angular momentum},
  pages={152--154},
  year={1990},
  publisher={IOP Publishing Ltd},
}

@article{forbes2016creation,
  title={Creation and detection of optical modes with spatial light modulators},
  author={Forbes, Andrew and Dudley, Angela and McLaren, Melanie},
  journal={Advances in optics and photonics},
  volume={8},
  number={2},
  pages={200--227},
  year={2016},
  publisher={Optical Society of America},
  url={https://doi.org/10.1364/AOP.8.000200}
}

@article{rubano2019q,
  title={Q-plate technology: a progress review},
  author={Rubano, Andrea and Cardano, Filippo and Piccirillo, Bruno and Marrucci, Lorenzo},
  journal={Journal of the optical society of america B},
  volume={36},
  number={5},
  pages={D70--D87},
  year={2019},
  publisher={Optical Society of America},
  url={https://doi.org/10.1364/JOSAB.36.000D70}
}

@article{karimi2014generating,
  title={Generating optical orbital angular momentum at visible wavelengths using a plasmonic metasurface},
  author={Karimi, Ebrahim and Schulz, Sebastian A and De Leon, Israel and Qassim, Hammam and Upham, Jeremy and Boyd, Robert W},
  journal={Light: Science \& Applications},
  volume={3},
  number={5},
  pages={e167--e167},
  year={2014},
  publisher={Nature Publishing Group},
  url={https://doi.org/10.1038/lsa.2014.48}
}

@article{beresna2011radially,
  title={Radially polarized optical vortex converter created by femtosecond laser nanostructuring of glass},
  author={Beresna, Martynas and Gecevi{\v{c}}ius, Mindaugas and Kazansky, Peter G and Gertus, Titas},
  journal={Applied Physics Letters},
  volume={98},
  number={20},
  year={2011},
  publisher={AIP Publishing},
  url={https://doi.org/10.1063/1.3590716}
}

@article{labroille2014efficient,
  title={Efficient and mode selective spatial mode multiplexer based on multi-plane light conversion},
  author={Labroille, Guillaume and Denolle, Bertrand and Jian, Pu and Genevaux, Philippe and Treps, Nicolas and Morizur, Jean-Fran{\c{c}}ois},
  journal={Optics express},
  volume={22},
  number={13},
  pages={15599--15607},
  year={2014},
  publisher={Optical Society of America},
  url={https://doi.org/10.1364/OE.22.015599}
}

@article{fontaine2019laguerre,
  title={Laguerre-Gaussian mode sorter},
  author={Fontaine, Nicolas K and Ryf, Roland and Chen, Haoshuo and Neilson, David T and Kim, Kwangwoong and Carpenter, Joel},
  journal={Nature communications},
  volume={10},
  number={1},
  pages={1865},
  year={2019},
  publisher={Nature Publishing Group UK London},
  url={https://doi.org/10.1038/s41467-019-09840-4}
}

@inproceedings{fontaine2019digital,
  title={Digital turbulence compensation of free space optical link with multimode optical amplifier},
  author={Fontaine, Nicolas K and Ryf, Roland and Zhang, Yuanhang and Alvarado-Zacarias, Juan Carlos and van der Heide, Sjoerd and Mazur, Mikael and Huang, Hanzi and Chen, Haoshuo and Amezcua-Correa, Rodrigo and Li, Guifang and others},
  booktitle={45th European Conference on Optical Communication (ECOC 2019)},
  pages={1--4},
  year={2019},
  organization={IET},
  url={https://doi.org/10.1049/cp.2019.1015}
}

@article{mounaix2020time,
  title={Time reversed optical waves by arbitrary vector spatiotemporal field generation},
  author={Mounaix, Mickael and Fontaine, Nicolas K and Neilson, David T and Ryf, Roland and Chen, Haoshuo and Alvarado-Zacarias, Juan Carlos and Carpenter, Joel},
  journal={Nature communications},
  volume={11},
  number={1},
  pages={5813},
  year={2020},
  publisher={Nature Publishing Group UK London},
  url={https://doi.org/10.1038/s41467-020-19601-3}
}

@article{boucher2020spatial,
  title={Spatial optical mode demultiplexing as a practical tool for optimal transverse distance estimation},
  author={Boucher, Pauline and Fabre, Claude and Labroille, Guillaume and Treps, Nicolas},
  journal={Optica},
  volume={7},
  number={11},
  pages={1621--1626},
  year={2020},
  publisher={Optical Society of America},
  url={https://doi.org/10.1364/OPTICA.404746}
}

@article{brandt2020high,
  title={High-dimensional quantum gates using full-field spatial modes of photons},
  author={Brandt, Florian and Hiekkam{\"a}ki, Markus and Bouchard, Fr{\'e}d{\'e}ric and Huber, Marcus and Fickler, Robert},
  journal={Optica},
  volume={7},
  number={2},
  pages={98--107},
  year={2020},
  publisher={Optical Society of America},
  url={https://doi.org/10.1364/OPTICA.375875}
}

@article{lib2022processing,
  title={Processing entangled photons in high dimensions with a programmable light converter},
  author={Lib, Ohad and Sulimany, Kfir and Bromberg, Yaron},
  journal={Physical Review Applied},
  volume={18},
  number={1},
  pages={014063},
  year={2022},
  publisher={APS},
  url={https://doi.org/10.1103/PhysRevApplied.18.014063}
}

@article{hiekkamaki2021high,
  title={High-dimensional two-photon interference effects in spatial modes},
  author={Hiekkam{\"a}ki, Markus and Fickler, Robert},
  journal={Physical Review Letters},
  volume={126},
  number={12},
  pages={123601},
  year={2021},
  publisher={APS},
  url={https://doi.org/10.1103/PhysRevLett.126.123601}
}

@article{wang2025ultracompact,
  title={Ultracompact 3D integrated photonic chip for high-fidelity high-dimensional quantum gates},
  author={Wang, Kangrui and Lyu, Dawei and Cai, Chengkun and Fu, Tianhao and Wang, Jue and Wang, Qianke and Liu, Jun and Wang, Jian},
  journal={Science Advances},
  volume={11},
  number={27},
  pages={eadv5718},
  year={2025},
  publisher={American Association for the Advancement of Science},
  url={https://doi.org/10.1126/sciadv.adv5718}
}

@article{lib2025high,
  title={High-dimensional quantum key distribution using a multi-plane light converter},
  author={Lib, Ohad and Sulimany, Kfir and Ara{\'u}jo, Mateus and Ben-Or, Michael and Bromberg, Yaron},
  journal={Optica Quantum},
  volume={3},
  number={2},
  pages={182--188},
  year={2025},
  publisher={Optica Publishing Group},
  url={https://doi.org/10.1364/OPTICAQ.531472}
}

@article{soma2025complete,
  title={Complete vectorial optical mode converter using multi-layer metasurface},
  author={Soma, Go and Komatsu, Kento and Nakano, Yoshiaki and Tanemura, Takuo},
  journal={Nature Communications},
  volume={16},
  number={1},
  pages={7744},
  year={2025},
  publisher={Nature Publishing Group UK London},
  url={https://doi.org/10.1038/s41467-025-62401-w}
}

@article{lin2018all,
  title={All-optical machine learning using diffractive deep neural networks},
  author={Lin, Xing and Rivenson, Yair and Yardimci, Nezih T and Veli, Muhammed and Luo, Yi and Jarrahi, Mona and Ozcan, Aydogan},
  journal={Science},
  volume={361},
  number={6406},
  pages={1004--1008},
  year={2018},
  publisher={American Association for the Advancement of Science},
  url={https://doi.org/10.1126/science.aat8084}
}

@article{hashimoto2005optical,
  title={Optical circuit design based on a wavefront-matching method},
  author={Hashimoto, T and Saida, T and Ogawa, I and Kohtoku, M and Shibata, Tomohiro and Takahashi, Hiroshi},
  journal={Optics letters},
  volume={30},
  number={19},
  pages={2620--2622},
  year={2005},
  publisher={Optica Publishing Group},
  url={https://doi.org/10.1364/OL.30.002620}
}

@article{sakakura2020ultralow,
  title={Ultralow-loss geometric phase and polarization shaping by ultrafast laser writing in silica glass},
  author={Sakakura, Masaaki and Lei, Yuhao and Wang, Lei and Yu, Yan-Hao and Kazansky, Peter G},
  journal={Light: Science \& Applications},
  volume={9},
  number={1},
  pages={15},
  year={2020},
  publisher={Nature Publishing Group UK London},
  url={https://doi.org/10.1038/s41377-020-0250-y}
}

@article{slussarenko2016guiding,
  title={Guiding light via geometric phases},
  author={Slussarenko, Sergei and Alberucci, Alessandro and Jisha, Chandroth P and Piccirillo, Bruno and Santamato, Enrico and Assanto, Gaetano and Marrucci, Lorenzo},
  journal={Nature Photonics},
  volume={10},
  number={9},
  pages={571--575},
  year={2016},
  publisher={Nature Publishing Group UK London},
  url={https://doi.org/10.1038/nphoton.2016.138}
}

@article{jisha2021geometric,
  title={Geometric phase in optics: from wavefront manipulation to waveguiding},
  author={Jisha, Chandroth Pannian and Nolte, Stefan and Alberucci, Alessandro},
  journal={Laser \& Photonics Reviews},
  volume={15},
  number={10},
  pages={2100003},
  year={2021},
  publisher={Wiley Online Library},
  url={https://doi.org/10.1002/lpor.202100003}
}

@article{beresna2014ultrafast,
  title={Ultrafast laser direct writing and nanostructuring in transparent materials},
  author={Beresna, Martynas and Gecevi{\v{c}}ius, Mindaugas and Kazansky, Peter G},
  journal={Advances in Optics and Photonics},
  volume={6},
  number={3},
  pages={293--339},
  year={2014},
  publisher={Optical Society of America},
  url={https://doi.org/10.1364/AOP.6.000293}
}

@article{verrier2011off,
  title={Off-axis digital hologram reconstruction: some practical considerations},
  author={Verrier, Nicolas and Atlan, Michael},
  journal={Applied optics},
  volume={50},
  number={34},
  pages={H136--H146},
  year={2011},
  publisher={Optica Publishing Group},
  url={https://doi.org/10.1364/AO.50.00H136}
}

@article{chavilkkadan2025purely,
  title={Purely geometric spin--orbit Laguerre--Gauss waveplates from 3D laser-nanostructured silica glass},
  author={Chavilkkadan, Mufeeduzaman and Shevtsov, Sergei and Kazansky, Peter and Brasselet, Etienne},
  journal={APL Photonics},
  volume={10},
  number={4},
  year={2025},
  publisher={AIP Publishing},
  url={https://doi.org/10.1063/5.0255787}
}

@article{beijersbergen1993astigmatic,
  title={Astigmatic laser mode converters and transfer of orbital angular momentum},
  author={Beijersbergen, Marco W and Allen, Les and Van der Veen, HELO and Woerdman, JP},
  journal={Optics Communications},
  volume={96},
  number={1-3},
  pages={123--132},
  year={1993},
  publisher={Elsevier},
  url={https://doi.org/10.1016/0030-4018(93)90535-D}
}

@article{butaite2026miniaturised,
  title={Miniaturised multi-plane light converters via laser-written geometric phase holograms},
  author={B{\=u}tait{\.e}, Un{\.e} G and Beresna, Martynas and Phillips, David B},
  journal={arXiv preprint arXiv:2602.07222},
  year={2026},
  url={https://doi.org/10.48550/arXiv.2602.07222}
}

@article{bouchard2019quantum,
  title={Quantum process tomography of a high-dimensional quantum communication channel},
  author={Bouchard, Fr{\'e}d{\'e}ric and Hufnagel, Felix and Koutn{\`y}, Dominik and Abbas, Aazad and Sit, Alicia and Heshami, Khabat and Fickler, Robert and Karimi, Ebrahim},
  journal={Quantum},
  volume={3},
  pages={138},
  year={2019},
  publisher={Verein zur F{\"o}rderung des Open Access Publizierens in den Quantenwissenschaften},
  url={	https://doi.org/10.22331/q-2019-05-06-138}
}

@article{zhang2025topological,
  title={Topological protection degrees of optical skyrmions and their electrical control},
  author={Zhang, Zan and Xie, Xi and Zhuang, Chuhong and Wu, Binyu and Liu, Zihan and Wu, Baoyun and Mihalache, Dumitru and Shen, Yijie and Deng, Dongmei},
  journal={Photonics Research},
  volume={13},
  number={9},
  pages={B1--B11},
  year={2025},
  publisher={Chinese Laser Press and Optica Publishing Group},
  url={https://doi.org/10.1364/PRJ.569522}
}

@article{ornelas2025topological,
  title={Topological rejection of noise by quantum skyrmions},
  author={Ornelas, Pedro and Nape, Isaac and de Mello Koch, Robert and Forbes, Andrew},
  journal={Nature Communications},
  volume={16},
  number={1},
  pages={2934},
  year={2025},
  publisher={Nature Publishing Group UK London},
  url={https://doi.org/10.1038/s41467-025-58232-4}
}

@article{pushkina2021superresolution,
  title={Superresolution linear optical imaging in the far field},
  author={Pushkina, AA and Maltese, G and Costa-Filho, JI and Patel, P and Lvovsky, AI},
  journal={Physical review letters},
  volume={127},
  number={25},
  pages={253602},
  year={2021},
  publisher={APS},
  url={https://doi.org/10.1103/PhysRevLett.127.253602}
}

@article{lib2025building,
  title={Building and aligning a 10-plane light converter},
  author={Lib, Ohad and Shekel, Ronen and Bromberg, Yaron},
  journal={Journal of Physics: Photonics},
  volume={7},
  number={3},
  pages={033001},
  year={2025},
  publisher={IOP Publishing},
  url={https://doi.org/10.1088/2515-7647/add06b}
}

@article{shimotsuma2003self,
  title={Self-organized nanogratings in glass irradiated by ultrashort light pulses},
  author={Shimotsuma, Yasuhiko and Kazansky, Peter G and Qiu, Jiarong and Hirao, Kazuoki},
  journal={Physical review letters},
  volume={91},
  number={24},
  pages={247405},
  year={2003},
  publisher={APS},
  url={https://doi.org/10.1103/PhysRevLett.91.247405}
}

@article{microsoft2026laser,
  title={Laser writing in glass for dense, fast and efficient archival data storage},
   author={{Microsoft Research Project Silica Team}},
  journal={Nature},
  volume={650},
  number={8102},
  pages={606--612},
  year={2026},
  publisher={Nature Publishing Group UK London},
  url={https://doi.org/10.1038/s41586-025-10042-w}
}

@article{zhang2025recent,
  title={Recent advances in femtosecond laser direct writing of three-dimensional periodic photonic structures in transparent materials},
  author={Zhang, Bin and Yan, Wenchao and Chen, Feng},
  journal={Advanced Photonics},
  volume={7},
  number={3},
  pages={034002--034002},
  year={2025},
  publisher={Society of Photo-Optical Instrumentation Engineers},
  url={https://doi.org/10.1117/1.AP.7.3.034002}
}

@article{liu2025control,
  title={Control of optical skyrmionic textures via the Pancharatnam-Berry phase},
  author={Liu, Guang and Shi, Lujia and Liu, Yanzhe and Zeng, Xinji and Wang, Jinwen and Fu, Zhenbin and Zhang, Yingxin and Chen, Haixia and Wei, Dong},
  journal={Physical Review A},
  volume={112},
  number={5},
  pages={053507},
  year={2025},
  publisher={APS},
  url={https://doi.org/10.1103/frvz-v9qq}
}
% For visualization of modal leakage in the sorters, logarithmic crosstalk matrices were calculated as
%\begin{equation}
%C_{ij}^{(\mathrm{dB})} = 10 \log_{10}(100\,C_{ij}).
%\end{equation}
%The off-diagonal elements of the logarithmic crosstalk matrix define the intermodal crosstalk levels, while the average coupling loss relative to the target Gaussian modes was obtained from the mean of the diagonal elements.

%%%%%%%%%%%%%%
\clearpage

\begin{widetext}
\begin{center}
\Large\textbf{Supplementary Information: Volumetric processing of Structured Light Integrated in Glass}
\end{center}

%\section*{Supplementary Information: Volumetric processing of Structured Light Integrated in Glass}
\subsection*{Femtosecond laser writing setup and fabrication procedure}
The modulation planes were fabricated inside fused silica using a femtosecond laser direct writing technique, as described in the Methods section of the main text. The retardance induced in the material depends on the pulse density and the pulse energy delivered to the sample. In this work, the pulse density was fixed by setting the sample translation speed to 0.2~mm/s and the laser repetition rate to 250~kHz. The retardance was therefore controlled by adjusting the average laser power, which directly determines the pulse energy.
The average laser power was controlled during the writing process using an attenuator consisting of a half-wave plate (HWP) and a polarizing beam splitter (PBS), as shown in the detailed setup in Fig \ref{FigSupp1}.(a). 

To control the fast-axis orientation of the written birefringence, the polarization of the writing beam was set to be linear and dynamically modulated, i.e., it linear polarization rotated, using two electro-optic modulators (EOMs) driven by a programmable waveform generator, together with a half-wave plate (HWP) and a quarter-wave plate (QWP). Each modulation plane was written line by line, with the fast-axis orientation perpendicular to the laser polarization. The corresponding polarization-control waveforms were uploaded to the waveform generator prior to fabrication and applied during inscription.
For calibration purposes, several test samples containing 2 layers of nanogratings separated by 100 $\mu m$ were fabricated. 
The written test pattern of the nanogratings consisted of an area of 200 $\mu$m by 200 $\mu$m and follows an azimuthal gradient corresponding to an optical vortex with topological charge $\ell = 1$. 
This helical pattern provides a continuous variation of the fast-axis orientation across the sample, allowing convenient characterization of both the local retardance and the fast-axis orientation after fabrication. 
\begin{figure}[H]
    \centering
    \includegraphics[width=0.9\linewidth]{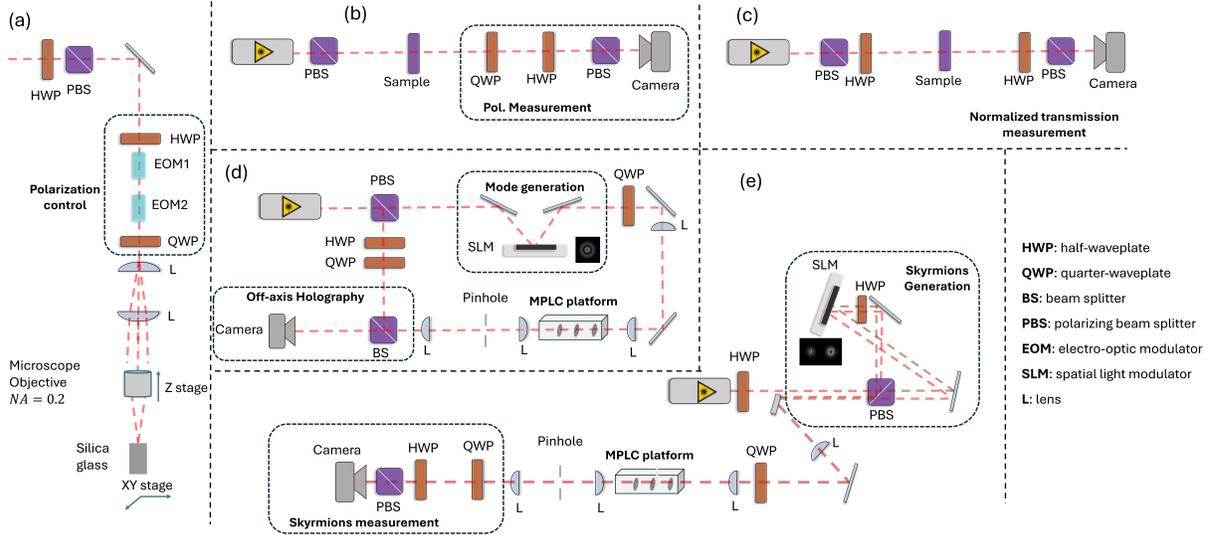}
    \caption{\textbf{Experimental setups used for fabrication and characterization of the volumetric MPLC platform.} (a) Schematic of the femtosecond direct laser-writing setup used to inscribe the birefringent modulation planes inside fused silica glass. (b) Experimental setup for polarization tomography, used to reconstruct the fast-axis orientations of the structured half-wave plates. (c) Setup used for normalized transmission measurements, where a normalized transmission of 1 corresponds to a retardance of $\pi$. (d) Experimental setup for MPLC characterization. The input mode is generated using a spatial light modulator (SLM), coupled into the MPLC glass device, and the resulting complex output field is reconstructed using off-axis holography. (e) Setup used to generate skyrmion beams with a Sagnac interferometer and couple them into the MPLC device. Polarization tomography is then performed to reconstruct the full Stokes-parameter distribution of the output skyrmion field.
 }
    \label{FigSupp1}
\end{figure}
The fabricated samples were subsequently characterized for 808 nm and 1550 nm wavelengths using the polarization tomography and retardance measurements described in the following sections.
\subsection*{Polarization tomography for fast-axis reconstruction}
The local orientation of the fast-axis of the fabricated structures was measured using polarization tomography. 
A horizontally polarized beam was expanded and then transmitted through the sample.
The resulting spatially-varying polarization state was analyzed using a standard Stokes polarimetry setup consisting of a (QWP), a (HWP), and a polarizing beam splitter (PBS), along with a camera for spatial resolution. 
The detailed setup is shown in Fig \ref{FigSupp1}.(b). 
By rotating the wave plates, six polarization projections were recorded corresponding to the horizontal (H), vertical (V), diagonal (D), anti-diagonal (A), right-circular (R), and left-circular (L) polarization bases.

From these intensity measurements, the Stokes parameters $(S_0,S_1,S_2,S_3)$ were calculated as
\[
S_0 = I_H + I_V, \qquad
S_1 = I_H - I_V,
\]
\[
S_2 = I_D - I_A, \qquad
S_3 = I_R - I_L,
\]
where $I_H$, $I_V$, $I_D$, $I_A$, $I_R$, and $I_L$ denote the measured intensities of the corresponding polarization states.

For a birefringent structure acting as a wave plate, the orientation of the fast-axis $\theta$ can be extracted from the linear polarization components of the Stokes vector. The fast-axis angle was calculated according to
\[
\theta = \frac{1}{2}\arctan\!\left(\frac{S_2}{S_1}\right).
\]

Since the arc-tangent function returns values in the interval $[-\pi/2,\pi/2]$, the resulting orientation map exhibits discontinuities. 

\begin{figure}[htb]
    \centering
    \includegraphics[width=0.9\linewidth]{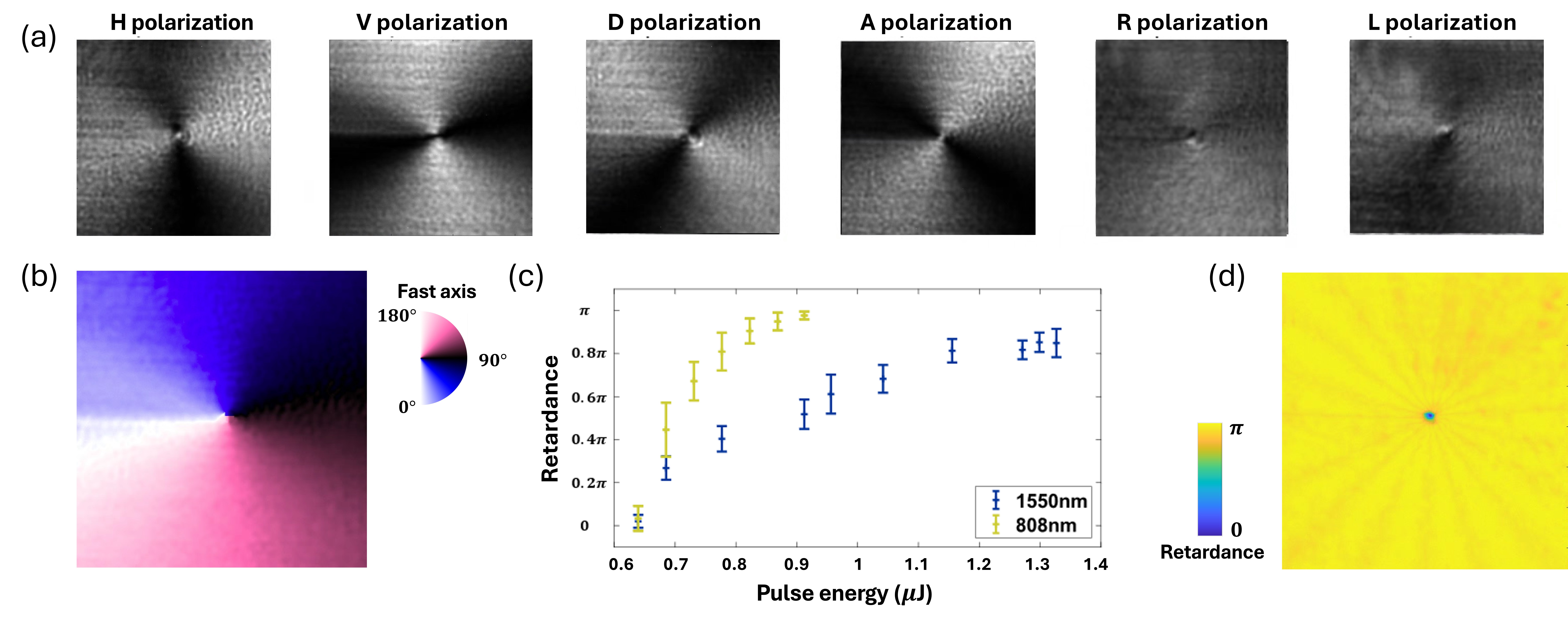}
\caption{\textbf{Polarization tomography and retardance characterization of a structured half-wave plate.}
(a) Measured intensity distributions corresponding to the six polarization projection bases (H, V, D, A, R, and L) used for polarization tomography of a structured half-wave plate with fast-axis orientation corresponding to a plate imprinting an OAM charge of $\ell = 1$.
(b) Reconstructed spatial distribution of the fast-axis orientation for the same structured half-wave plate.
(c) Retrieved spatial distribution of the birefringent retardance for the same sample, showing values close to the half-wave condition ($\pi$ retardance).
(d) Spatially averaged retardance as a function of femtosecond writing pulse energy at wavelengths of 808\,nm and 1550\,nm, extracted from polarization tomography measurements of samples fabricated with different pulse energies.
}
    \label{FigSupp2}
\end{figure}
To recover a continuous fast-axis distribution across the sample, a numerical phase-unwrapping procedure was applied, allowing the orientation to be reconstructed in the range $[0,\pi]$. 
This measurement provides the spatial distribution of the birefringent axis orientation written inside the glass and enables verification of the designed modulation patterns.
Fig \ref{FigSupp2}.(a) shows the six recorded polarization projection images (H, V, D, A, R, L). %Background artifacts visible in the images arise from reflections in the protective glass in front of the camera sensor. %Since these features remain fixed for all recorded polarization projections, they do not influence the polarization-tomography reconstruction or the retrieved fast-axis orientation and retardance maps. 
Fig \ref{FigSupp2}.(b) shows the retrieved axis-orientation for the same sample with HWP retardance. 

The same polarization-tomography procedure was used to retrieve the fast-axis orientation maps of the modulation planes shown in the main text. By slightly tilting the glass sample, the modulation planes become laterally separated in the image plane, allowing each mask to be reconstructed individually.
\subsection*{Retardance characterization}
The local retardance of the fabricated structures was characterized using a transmission measurement between two polarizers.
The sample was placed between a polarizer and an analyzer, each consisting of a half-wave plate (HWP) followed by a polarizing beam splitter (PBS). 
By rotating both wave plates simultaneously, the relative angle between the incident polarization and the optical axis of the birefringent structure could be varied. In principle, the same effect could be achieved by rotating the sample. However, since the transmitted intensity was recorded with a camera to retrieve pixel-resolved retardance values, the sample was kept fixed to ensure that each camera pixel corresponded to the same spatial location on the sample throughout the measurement.
The normalized transmitted intensity was recorded as a function of the angle $\alpha$ between the incident linear polarization and the fast-axis of the sample. 
For an ideal birefringent element with retardance $\delta$, the normalized transmission is given by
\[
T_n(\alpha)=\frac{1}{2}\sin^2(2\alpha)\left[1-\cos(\delta)\right].
\]

This relation produces a sinusoidal modulation of the transmitted intensity as the polarization angle is varied. 
The amplitude of this modulation depends on the local retardance $\delta$ of the sample. 
In particular, a maximum normalized transmission of $T_n^{\mathrm{max}}=1$ is obtained when the retardance reaches $\delta=\pi$, corresponding to half-wave plate behavior.

By measuring the transmission modulation across the fabricated structures, the spatial distribution of the induced retardance was determined. 
The retardance measurement setup is shown in \ref{FigSupp1}.(c).

\subsection*{Calibration of pulse energy for $\pi$ retardance}
To determine the pulse energy required to induce a retardance of $\pi$, a series of calibration samples imprinting an OAM of charge $\ell=1$ (as before) were fabricated in the same $z$-plane close to the surface of the glass. The samples are laterally separated by $100\,\mu\mathrm{m}$, and written with different pulse energies. The pulse energy was varied by adjusting the average laser power during the writing process. For each fabricated sample, the local retardance was measured using the characterization method described in the previous section. The retardance value was calculated for each pixel within the measurement area, and the average retardance was obtained by averaging over the full physical sample area of $200 \times 200\,\mu\mathrm{m}^2$.
Fig \ref{FigSupp2}.(c) shows the dependence of the averaged retardance on the applied pulse energy for both 808 nm and 1550nm. 
The retardance increases with increasing pulse energy due to the stronger laser-induced modification of the glass structure. 
At 808 nm, retardance values close to $\pi$ were achieved. However, at 1550 nm, higher pulse energies would be required to reach $\pi$ retardance, which exceeded the available laser power. 
To still achieve a full $\pi$-retardance, an additional layer of nanogratings could be manufactured. 
The error bars describe the standard deviation of the retardance values within this region. 
The retardance map of a sample that was fabricated using the optimal pulse energy for 808 nm (0.9 µJ) is shown Fig \ref{FigSupp2}.(d) and demonstrates that the retardance approaches $\pi$ across most of the structure, confirming that the chosen pulse energy produces the desired half-wave plate behavior.

\subsection*{Depth-dependent retardance calibration}
As described in the method section, two depth compensation strategies were used. In the first strategy, the writing beam was collimated before entering the microscope objective. As the focal position was translated to increasing depths inside the substrate, the focal spot broadened and the induced birefringent retardance decreased with depth when constant writing parameters were used. To compensate for this effect, the pulse energy was progressively increased for deeper planes, as shown in Fig \ref{FigSupp3}(a).
In the second strategy, the writing beam was made slightly divergent before entering the microscope objective. As the objective was translated to deeper positions inside the glass, the beam diameter at the entrance aperture increased accordingly, compensating for the depth-induced focal broadening. This approach enabled a nearly constant focal-spot size throughout the full writing range and resulted in a stable birefringent retardance, as shown in Fig \ref{FigSupp3}(b).
The effectiveness of the second approach was verified by measuring the evolution of the beam waist near the focus for two objective positions, $z_0$ and $z_1 = z_0 + 2.2$~mm, corresponding to approximately 3~cm difference in writing depth inside the substrate.  For each position, the beam waist was recorded by translating the camera together with the sample stage in steps of 0.8~mm along the propagation direction. The measured beam-waist evolution is shown in Fig \ref{FigSupp3}.(c). The beam-waist measurements show that larger beam sizes are observed at position $z_0$, which is closer to the microscope objective, compared to position $z_1$. This behavior confirms that the writing beam is divergent before entering the objective. 
\begin{figure}[htb]
    \centering
    \includegraphics[width=0.9\linewidth]{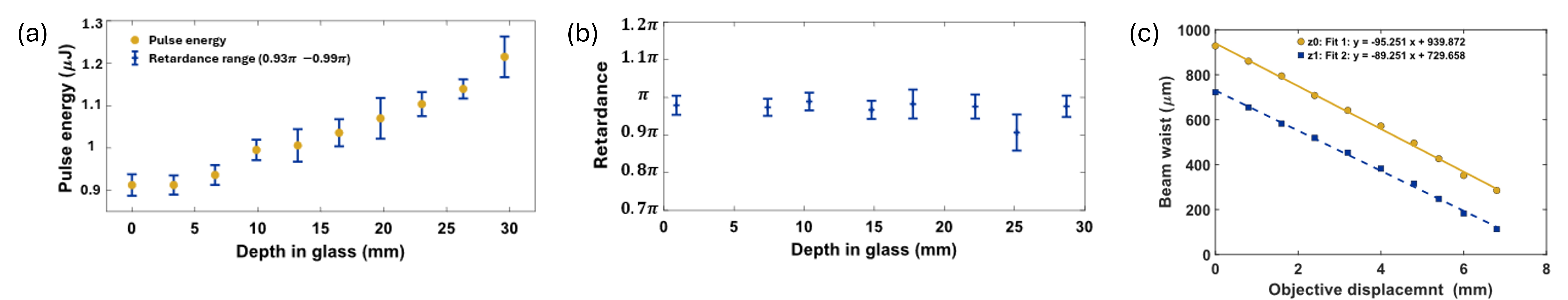}
\caption{\textbf{Depth-dependent calibration of femtosecond laser writing parameters.}
(a) Pulse energy required to obtain $\pi$ retardance versus writing depth inside fused silica. Error bars indicate the retardance range (0.93$\pi$–0.99$\pi$).
(b) Retardance at different depths using a fixed pulse energy of $0.9$ µJ, confirming near half-wave plate retardance across the writing range.
(c) Beam-waist measurements versus objective displacement at positions $z_0$ (yellow) and $z_1=z_0+2.2$ mm (blue).
}
    \label{FigSupp3}
\end{figure}

\subsection*{Synchronization-induced fabrication errors}
The main source of fabrication error in our system comes from imperfect synchronization between waveform generation (applied to the two EOMs) and translation-stage motion during laser writing. In the current implementation, both processes are controlled sequentially in \textsc{Matlab} without hardware-level triggering, which introduces timing offsets between adjacent written lines.
Although a fixed delay was introduced to compensate for this offset, the optimal delay varies between lines by approximately $\pm 5\%$, resulting in relative shifts between neighboring structures. This effect is clearly visible in periodic grating test patterns, where neighboring lines show visible shifts compared to the ideal grating pattern (see comparison in Fig.~\ref{FigSupp4}(a) and (b)).
This effect is still present for gratings with a single period (Fig.~\ref{FigSupp4}(c) and (d)). However, in this case the distortion is less pronounced and still allows good writing quality. For more complex masks, this effect increasingly reduces the performance of the MPLC.
Timing measurements of \textsc{Matlab}-controlled scan-line execution confirm fluctuations on the order of $\pm 5\%$, consistent with the observed spatial offsets. Therefore, Hardware-level synchronization is expected to significantly improve fabrication accuracy in future implementations.
\begin{figure}[H]
    \centering
    \includegraphics[width=0.5\linewidth]{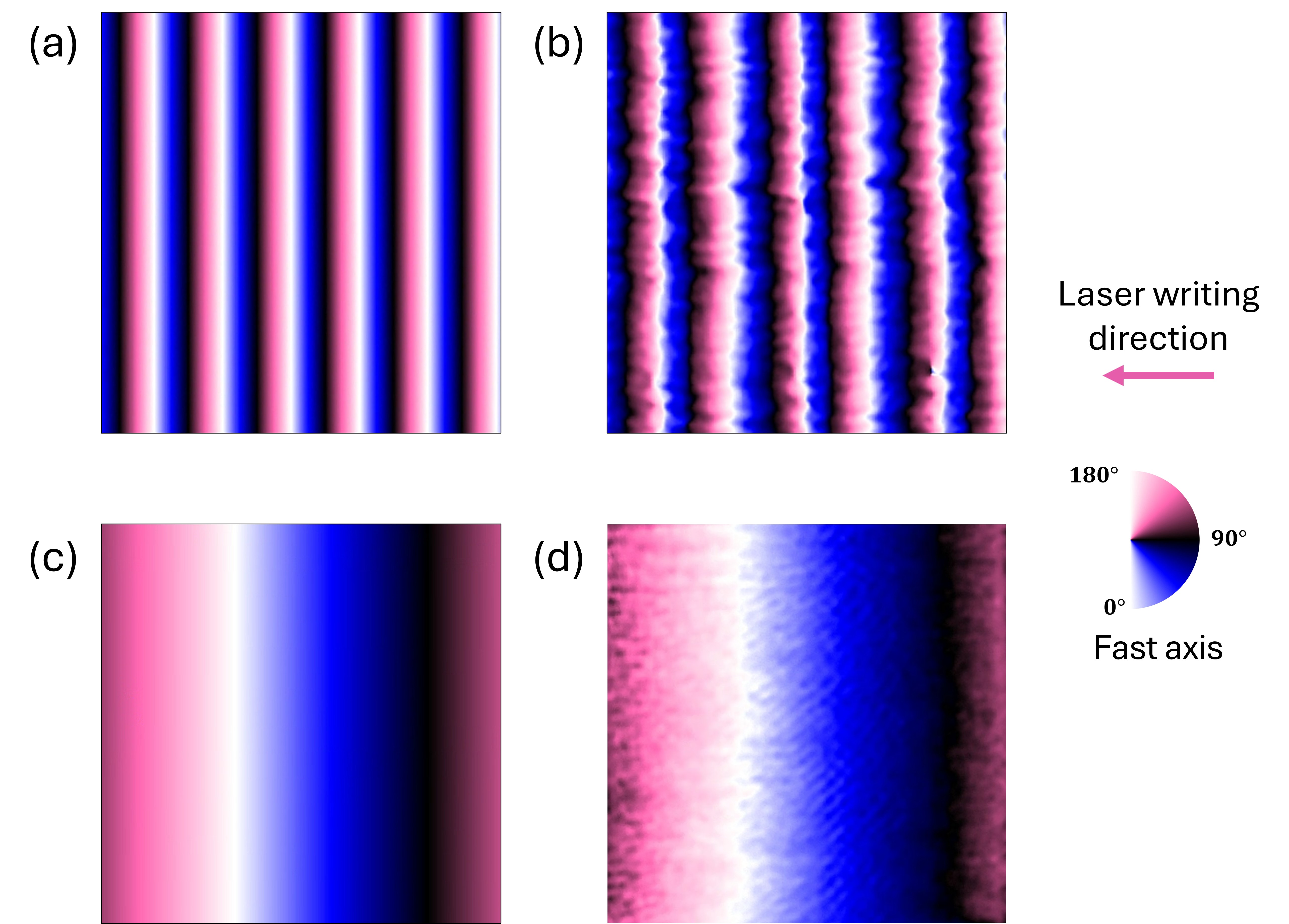}
\caption{\textbf{
Effect of synchronization errors between waveform generation and stage motion during laser writing.} 
(a) Ideal periodic fast-axis grating pattern. 
(b) Experimentally reconstructed periodic grating showing relative shifts between adjacent lines due to timing offsets along the writing direction (arrow). 
(c) Ideal single-period fast-axis pattern. 
(d) Experimentally reconstructed single-period pattern, where the distortion is still present but less pronounced. 
}

    \label{FigSupp4}
\end{figure}
\subsection*{Implemented high-dimensional unitary transformations}
The examples of high-dimensional spatial unitary transformations that were experimentally implemented using the MPLC platform are defined below.
The three-dimensional Hadamard transformation corresponds to a discrete Fourier transform in a three-dimensional modal subspace and is defined as
\begin{equation}
H_3 =
\frac{1}{\sqrt{3}}
\sum_{j,k=0}^{2}
e^{i\frac{2\pi}{3}jk}
|j\rangle \langle k|,
\end{equation}

with matrix representation
\begin{equation}
H_3 =
\frac{1}{\sqrt{3}}
\begin{pmatrix}
1 & 1 & 1 \\
1 & e^{i\frac{2\pi}{3}} & e^{i\frac{4\pi}{3}} \\
1 & e^{i\frac{4\pi}{3}} & e^{i\frac{2\pi}{3}}
\end{pmatrix}.
\end{equation}

The four-dimensional Hadamard transformation matrix is defined as
\begin{equation}
H_4 =
\frac{1}{2}
\begin{pmatrix}
1 & 1 & 1 & 1 \\
1 & i & -1 & -i \\
1 & -1 & 1 & -1 \\
1 & -i & -1 & i
\end{pmatrix}.
\end{equation}

The five-dimensional cyclic shift ($X$) gate performs a permutation of the modal basis states according to
\begin{equation}
X_5 =
\sum_{k=0}^{4}
|k+1 \; \mathrm{mod}\; 5\rangle \langle k|,
\end{equation}

with matrix representation

\begin{equation}
X_5 =
\begin{pmatrix}
0 & 1 & 0 & 0 & 0 \\
0 & 0 & 1 & 0 & 0 \\
0 & 0 & 0 & 1 & 0 \\
0 & 0 & 0 & 0 & 1 \\
1 & 0 & 0 & 0 & 0
\end{pmatrix}.
\end{equation}

In addition, polarization-controlled spatial transformations were implemented using the polarization-dependent geometric phase of the modulation planes. 
In this configuration, the circular polarization state acts as the control degree of freedom, while the spatial mode represents the target subspace.

A polarization-controlled NOT (cNOT) transformation acting on the spatial-mode subspace is defined as
\begin{equation}
U_{\mathrm{cNOT}} =
|L\rangle\!\langle L| \otimes I
+
|R\rangle\!\langle R| \otimes X ,
\end{equation}

with matrix representation
\begin{equation}
U_{\mathrm{CNOT}} =
\begin{pmatrix}
I & 0 \\
0 & X
\end{pmatrix}.
\end{equation}

In addition, a polarization-controlled multiplexing of two different spatial transformation was implemented, in which the operator $X$ is applied to left-circular polarization and the operator $H$ is applied to right-circular polarization. 
This transformation is defined as
\begin{equation}
U_{XH} =
|L\rangle\!\langle L| \otimes X
+
|R\rangle\!\langle R| \otimes H ,
\end{equation}

with matrix representation
\begin{equation}
U_{XH} =
\begin{pmatrix}
X & 0 \\
0 & H
\end{pmatrix}.
\end{equation}

\subsection*{Retrieved MPLC masks and additional experimental results}
In addition to the utilized modulation masks corresponding to the 3D Hadamard gate shown in the main text, the masks corresponding to the other demonstrated transformations are shown in Fig \ref{FigSupp5}. 

\begin{figure}[b]
    \centering
    \includegraphics[width=0.9\linewidth]{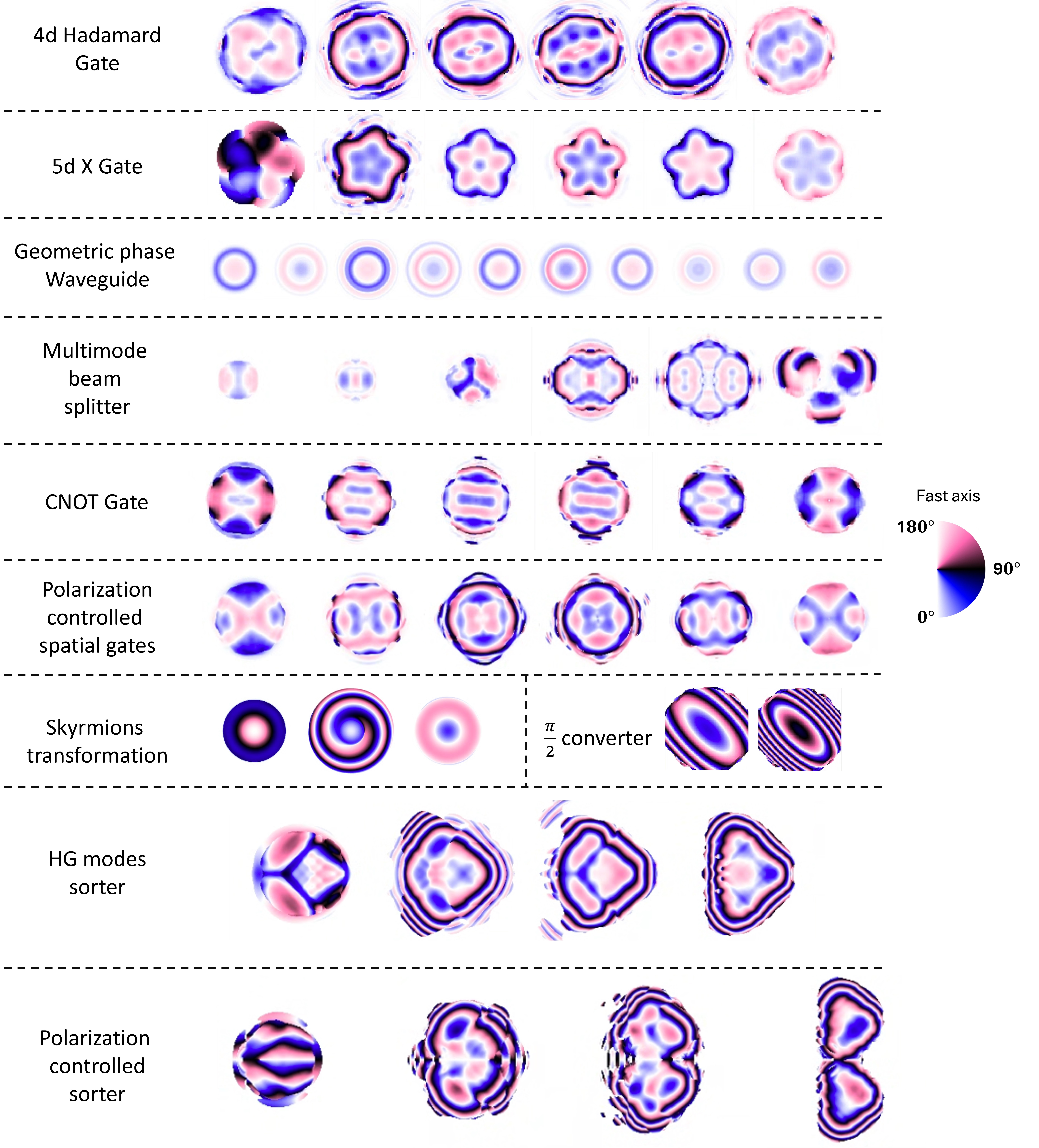}
\caption{Fast-axis orientation distributions of the fabricated modulation planes used in the main demonstrations.}

    \label{FigSupp5}
\end{figure}

Table~S1 summarizes the main MPLC parameters for each device, including the mask size, number of modulation planes, writing-depth compensation strategy, pixel size, input beam waist, output beam waist, and the additional free-space propagation distance after the MPLC structure.

\begin{table*}[htb]
\centering
\caption{\textbf{Summary of fabrication and optical parameters of the demonstrated MPLC devices.}}
\label{Table_MPLC_parameters}
\setlength{\tabcolsep}{4pt}
\footnotesize
\resizebox{\textwidth}{!}{
\begin{tabular}{c c c c c c c c c}
\hline
MPLC & $\lambda$ & Mask size & Number of planes & Distance between & Pixel size & Input beam & Output beam & Propagation length \\
 & (nm) & ($\mu$m $\times$ $\mu$m) &  & planes (mm) & ($\mu$m) & waist ($\mu$m) & waist ($\mu$m) & after MPLC (mm) \\
\hline

3D H gate & 808 & 500 $\times$ 500 & 6 & 5.6 & 8 & 50 & 50 & 0 \\
4D H gate & 808 & 500 $\times$ 500 & 6 & 5.6 & 8 & 50 & 50 & 0 \\
5D X gate & 808 & 500 $\times$ 500 & 6 & 5.6 & 8 & 50 & 50 & 0 \\
$\pi/2$ converter & 808 & 500 $\times$ 500 & 2 & 8 & 8 & 70 & 70 & 0 \\
Multimode beam splitter & 808 & 500 $\times$ 500 & 6 & 5.6 & 8 & 50 & 50 & 0 \\
Waveguide & 808 & 500 $\times$ 500 & 10 & 3.1 & 8 & 50 & 50 & 0 \\
cNOT gate & 808 & 500 $\times$ 500 & 6 & 5.6 & 8 & 70 & 70 & 0 \\
Polarization controlled X-H-gate & 808 & 500 $\times$ 500 & 6 & 5.6 & 2 & 60 & 60 & 0 \\
Skyrmion transformation & 808 & 500 $\times$ 500 & 3 & 11.5 & 2 & 60 & 60 & 0 \\
Scalar HG mode sorter & 1550 & 700 $\times$ 700 & 4 & 7.3 & 2 & 70 & 18 & 4 \\
Polarization controlled HG mode sorter  2 & 1550 & 700 $\times$ 700 & 4 & 7.3 & 2 & 40 & 15 & 4 \\

\hline
\end{tabular}
}
\end{table*}

Fig \ref{FigSupp6} shows the simulated output fields together with the experimentally reconstructed ones. 
The experimental beam profiles were retrieved using off-axis digital holography. 
In this method, the output beam interferes with a tilted reference beam, allowing reconstruction of both the amplitude and phase of the optical field from a single camera exposure. 
A schematic of the off-axis holography setup used in these measurements is shown in Fig \ref{FigSupp1}(d).
The optical losses of the fabricated MPLC devices were not measured directly for each structure. 
However, as indicated in the main text, at 808\,nm the average transmission per modulation plane was found to be approximately 89\%, corresponding to a loss of around 11\% per plane. 
Assuming similar transmission across the full device and accounting for multiple modulation planes as well as interface reflections from the uncoated glass substrate, the total transmission of the MPLC structures is estimated to be on the order of $\sim 45\%$. 
Additional losses arise from spatial filtering in the imaging system used after the MPLC during beam reconstruction and mode analysis. 
Taking these contributions into account, the overall transmission of the experimental system is estimated to be in the range of approximately 30--40\%.

\begin{figure}[b]
    \centering
    \includegraphics[width=0.9\linewidth]{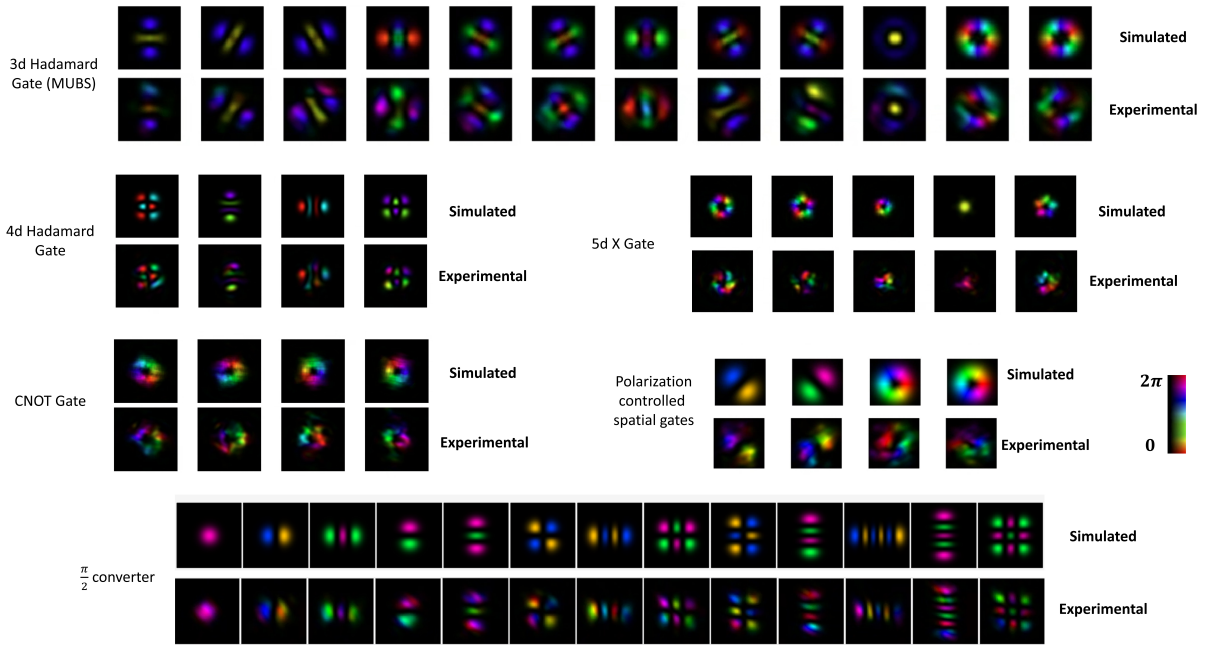}
    \caption{\textbf{Comparison between experimentally reconstructed and theoretical complex optical fields for the main demonstrations.}}
    \label{FigSupp6}
\end{figure}

For the mode-sorting devices operating at 1550\,nm, the optical losses are expected to differ from those measured at 808\,nm due to the non-ideal birefringent retardance of the laser-written structures at this wavelength. While a direct per-plane transmission measurement was measured to be 94\%, the overall transmission of the sorter is expected to be of a comparable order of magnitude to the values estimated for modulations at 808\,nm. 
A more precise determination would require dedicated wavelength-dependent loss characterization, which is beyond the current study but will be investigated in the future, focusing on telecom wavelength exclusively. 
Instead, the performance of the sorter at 1550\,nm is quantified through the experimentally reconstructed modal crosstalk matrices and the corresponding coupling losses to Gaussian output modes, which are reported in detail in Table S2.

\begin{table}
\centering
\caption{\textbf{Performance metrics of HG mode sorters at 1550\,nm.}}
\label{Table_sorter_performance}

\setlength{\tabcolsep}{6pt}
\footnotesize

\begin{tabular}{lcc}
\hline
 & Scalar HG mode sorter & Polarization controlled HG mode sorter\\
\hline

Diagonal overlaps (\%) & 0.39 to 0.71 & 0.13 to 0.33 \\
Off-diagonal overlaps (\%) & $5\times10^{-5}$ to 0.16 & $6\times10^{-6}$ to 0.036 \\
Mean diagonal overlap (\%) & 0.5609 & 0.24 \\
Mean off-diagonal overlap (\%) & 0.0093 & 0.004 \\
Coupling loss range (dB) & $-4.05$ to $-1.44$ & $-8.76$ to $-4.75$ \\
Intermodal crosstalk loss range (dB) & $-46.82$ to $-7.95$ & $-58.2$ to $-11.22$ \\
Average coupling loss (dB) & $-2.58$ & $-6.48$ \\
Average intermodal crosstalk loss (dB) & $-25.93$ & $-27.64$ \\

\hline
\end{tabular}
\end{table}

\subsection*{Polarization-controlled volumetric modulation and skyrmion transformations}
A distinctive feature of the presented volumetric MPLC platform is the use of polarization as an additional control degree of freedom. 
In particular, the birefringent geometric-phase modulation planes impose different phase shifts on right- and left-circular polarization components, enabling independent manipulation of the two polarization channels without requiring separate optical paths.
When expressed in the circular polarization basis, each birefringent modulation plane introduces opposite geometric phases for the two polarization components. 
As a result, a given phase mask produces conjugate wavefront transformations for the $|R\rangle$ and $|L\rangle$ components. 
For example, a lens-like phase profile generates opposite wavefront curvatures, such that one polarization component experiences an effective converging phase while the other experiences a diverging phase. 
During free-space propagation between successive modulation planes, this difference leads to a spatial separation of the two polarization components inside the MPLC volume.
This propagation-induced separation enables the wavefront-matching optimization algorithm to apply distinctly different transformations to the two polarization components while using a single birefringent modulation mask at each plane. 
As a result, different spatial operations can be implemented simultaneously for $|R\rangle$ and $|L\rangle$ within the same device. 

As an illustrative example, an MPLC was designed such that an input $\mathrm{LG}_{1,0}$ mode with right-circular polarization is transformed into the higher-order mode $\mathrm{LG}_{2,0}$, while a Gaussian input mode with left-circular polarization remains Gaussian. 
This specific polarization-controlled transformation corresponds to the manipulation of optical skyrmions that was presented in the main manuscript and discussed in the following in more detail.
Fig \ref{FigSupp7}(a) illustrates the underlying mechanism through numerical simulations of field propagation between successive modulation planes. 
The simulated intermediate planes clearly show the spatial separation of the two polarization components resulting from the opposite wavefront curvatures induced by the geometric-phase modulation. 
The separation enables independent optimization of the two polarization channels within the same volumetric MPLC structure.
As mentioned, this transformation describes the conversion between optical skyrmion states with different topological configurations.
Experimentally, the input skyrmion states were generated using a polarization-dependent Sagnac interferometer \cite{fickler2014quantum} combined with an imaging system, as shown in Fig.~\ref{FigSupp1}(e). 
After propagation through the MPLC device, the output Stokes vectors and polarization distributions were reconstructed using the polarization tomography setup described in Fig \ref{FigSupp1}(b) in detail. 
Fig \ref{FigSupp7}(b) presents a comparison between the target skyrmion polarization distributions and the experimentally reconstructed output polarization profiles obtained after the MPLC transformation, showing good agreement between theory and experiment.
The topology of the polarization field was quantified by calculating the corresponding skyrmion number defined as
\begin{equation}
N_{\mathrm{sk}}=
\frac{1}{4\pi}
\iint
\mathbf{n}(x,y)\cdot
\left(
\frac{\partial \mathbf{n}}{\partial x}
\times
\frac{\partial \mathbf{n}}{\partial y}
\right)
dx\,dy,
\end{equation}
where $\mathbf{n}(x,y) = (n_1,n_2,n_3)$ denotes the normalized Stokes vector obtained from the measured Stokes parameters $S_1(x,y)$, $S_2(x,y)$, and $S_3(x,y)$ as
\[
\mathbf{n}(x,y)=
\frac{\big(S_1(x,y),\,S_2(x,y),\,S_3(x,y)\big)}
{\sqrt{S_1^2(x,y)+S_2^2(x,y)+S_3^2(x,y)}}.
\]
Here, $x$ and $y$ denote the transverse spatial coordinates.
The integrand corresponds to the local skyrmion density, whose integration over the beam cross-section yields the total number of times the polarization field wraps the Poincaré sphere.
In practice, the integration was restricted to the spatial area of the beam where enough intensity was measured to suppress noise effects. 
A threshold equal to $5\%$ of the maximum intensity was used to define the area from which we evaluated the skyrmion number. 
The agreement between the target and experimentally reconstructed skyrmion numbers confirms the fidelity of the implemented polarization-topology transformations.

\begin{figure}
    \centering
    \includegraphics[width=0.5\linewidth]{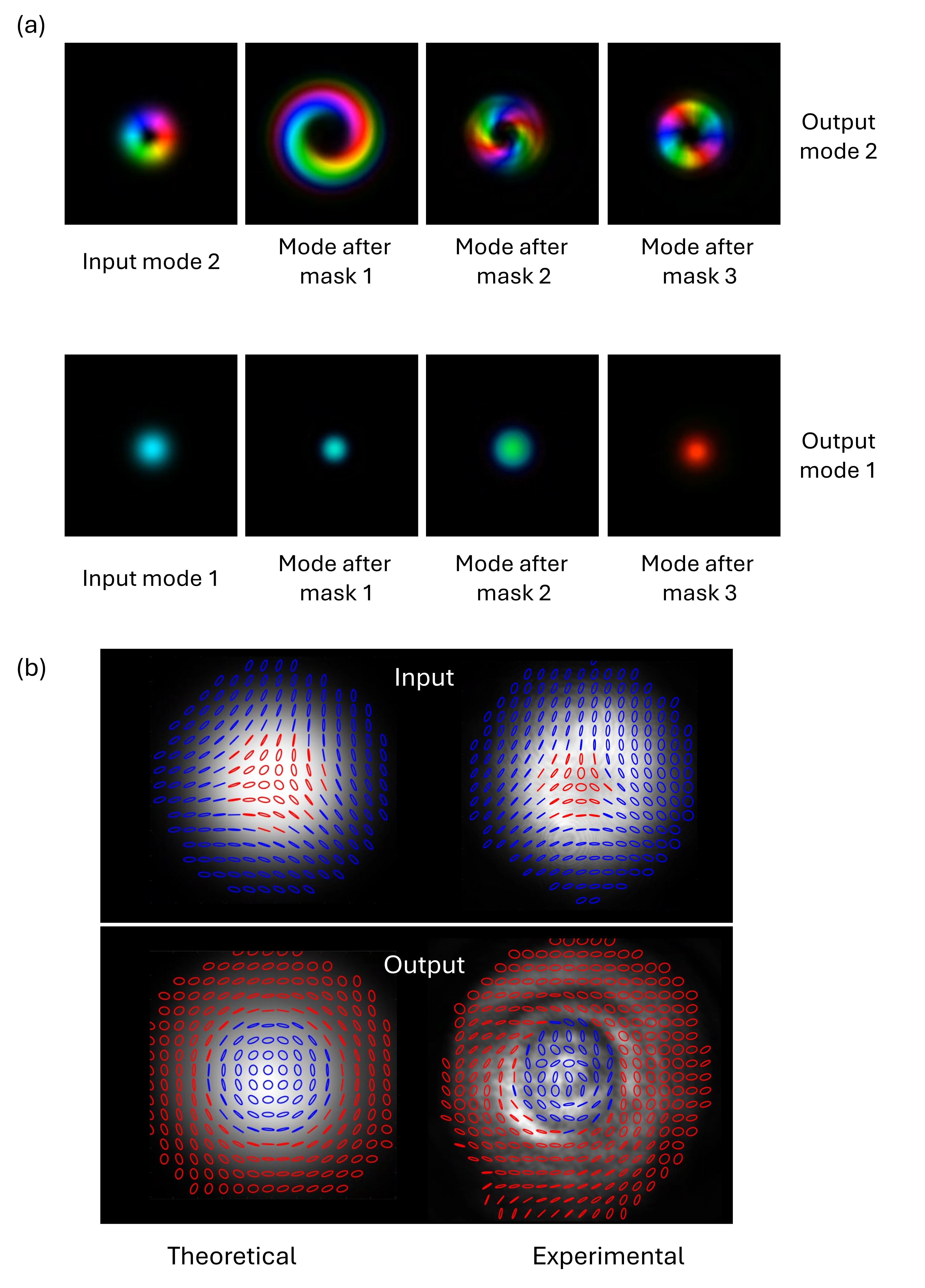}
\caption{\textbf{Skyrmion transformation using the volumetric MPLC device.}
\textbf{a)} Propagation of the two input modes through the MPLC device designed for skyrmion generation. The complex optical field (amplitude and phase) of each mode is shown after the modulation planes, illustrating the step-by-step evolution of the fields from the input to the final output plane.
\textbf{b)} Comparison between the input skyrmion field and the transformed output skyrmion field, shown for both theoretical and experimental reconstructions.
}
    \label{FigSupp7}
\end{figure}

\subsection*{Logarithmic representation of the crosstalk matrices at telecom wavelength}
To better visualize weak intermodal leakage in the telecom-wavelength mode-sorting experiments, the measured crosstalk matrices are additionally represented on a logarithmic (dB) scale in Fig \ref{FigSupp8}.(a) and Fig \ref{FigSupp8}.(b), for the scalar 15-mode HG sorter and the polarization-resolved 12-mode sorter, respectively.

\begin{figure}
    \centering
    \includegraphics[width=0.5\linewidth]{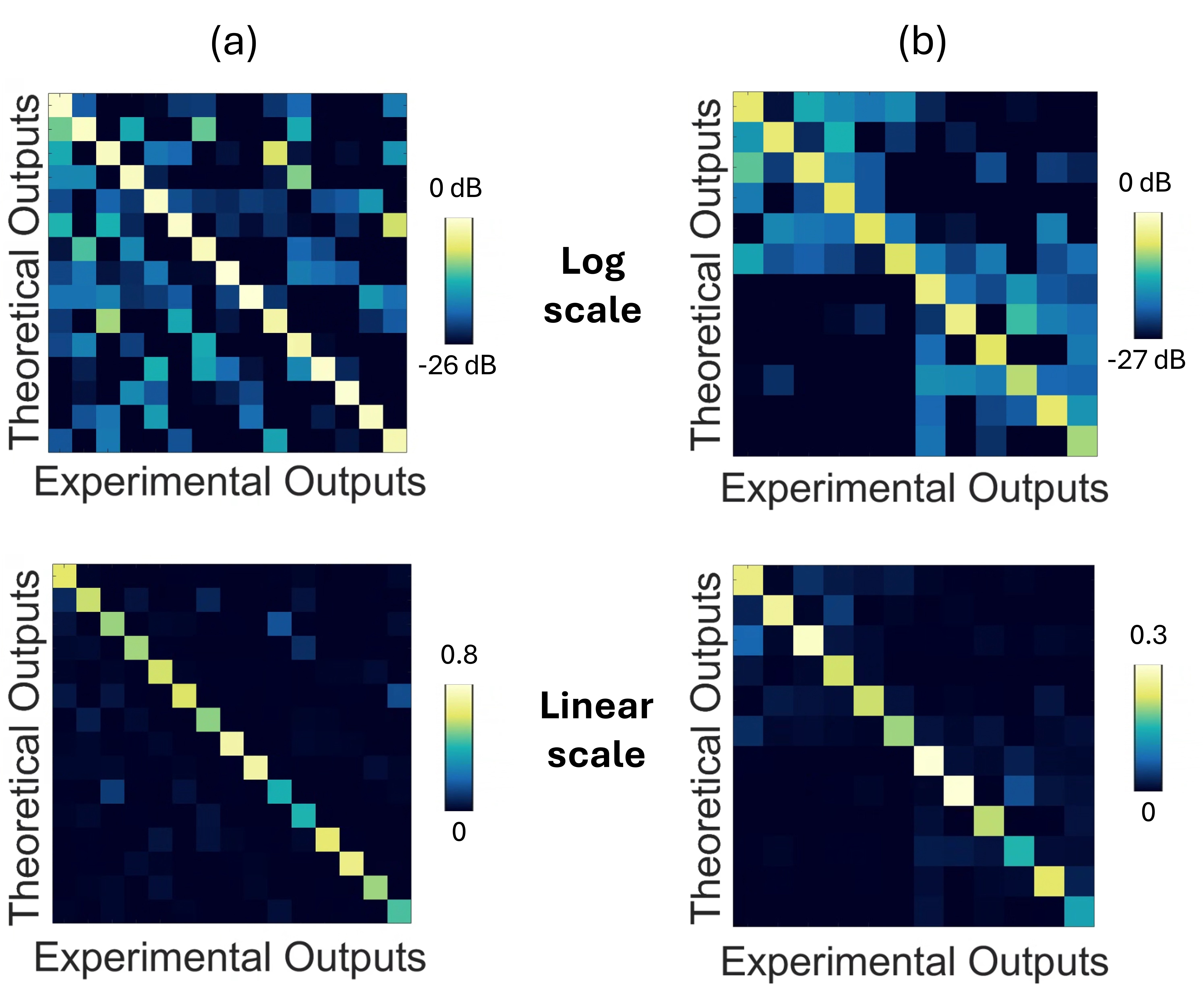}
\caption{\textbf{Linear and logarithmic representations of the measured crosstalk matrices at telecom wavelength}
(a) Crosstalk matrix of the 15-mode scalar Hermite–Gaussian (HG) mode sorter shown in logarithmic scale (top) and linear scale (bottom).
(b) Crosstalk matrix of the polarization-resolved HG mode sorter shown in logarithmic scale (top) and linear scale (bottom). For both sorters, the color scale is limited to the average off-diagonal crosstalk to visualize the relevant intermodal leakage, while the full numerical dataset is used for quantitative analysis.
}
    \label{FigSupp8}
\end{figure}
\end{widetext}

%\bibliography{references}

\end{document}